\documentclass[aps,twocolumn,showpacs,preprintnumbers,amsmath,amssymb]{revtex4}
\usepackage{epsf}
\usepackage{ulem} %% for strike-through
\usepackage[usenames]{color}

\newcommand{\MeV}{\textrm{ MeV}}
\newcommand{\be}{\begin{equation}}
\newcommand{\ee}{\end{equation}}
\newcommand{\ba}{\begin{eqnarray}}
\newcommand{\ea}{\end{eqnarray}}
\newcommand{\nn}{\nonumber}
\newcommand{\eee}{\epsilon}

\newcommand{\non}{\nonumber\\}

\newcommand{\Mmass}{m_{i}}
\newcommand{\Bmass}{M_{i}}
\newcommand{\Imass}{M_{N}}
\newcommand{\Fmass}{M_{N^{*}}}
\newcommand{\fpi}{f}
\newcommand{\prop}[2]{\, i \frac{#1}{#2+i\epsilon}}
\newcommand{\pini}{p_{i}}
\newcommand{\pfin}{P}
\newcommand{\pislash}{p_{i} \hspace{-8pt}/ \hspace{4pt}}
\newcommand{\Amp}{{\cal M}}

\newcommand{\Pslash}{{P\hspace{-7pt}/ \hspace{3pt}}}
\newcommand{\kslash}{{k\hspace{-5pt}/}}
\newcommand{\qslash}{{q\hspace{-5pt}/}}
\newcommand{\eslash}{{\epsilon \hspace{-4pt}/}}

\newcommand{\Aslash}{{A\hspace{-5.5pt}/}}

\newcommand{\intq}{\int \frac{d^{4}q}{(2\pi)^{4}}}
\newcommand{\intqd}{\int \frac{d^{d}q}{(2\pi)^{d}}}
\newcommand{\textfrac}[2]{\textstyle{\frac{#1}{#2}}}

\begin{document}

\preprint{YITP-07-83}

\title{Transition form factors of the $N^*(1535)$ as a dynamically generated
resonance }

\author{
D. Jido$^1$, M. D\"oring$^2$ and E.~Oset$^3$}
\affiliation{
$^1$ Yukawa Institute for Theoretical Physics, Kyoto University, Kyoto, 606-8502, Japan \\
$^2$ {Institut f\"ur Kernphysik, Forschungszentrum J\"ulich GmbH, 52425 J\"ulich, Germany}\\ 
$^3$ Departamento de F\'{\i}sica Te\'orica and IFIC,
Centro Mixto Universidad de Valencia-CSIC,
Institutos de Investigaci\'on de Paterna, Aptdo. 22085, 46071 Valencia, Spain
}

\date{\today}

\begin{abstract} 
We discuss how electromagnetic properties provide useful tests of the nature of resonances, and we study these properties for the $N^*(1535)$ which appears dynamically generated from the strong interaction of mesons and baryons. Within this coupled channel chiral unitary approach,
we evaluate the $A_{1/2}$  and $S_{1/2}$ helicity amplitudes as a function
 of $Q^2$ for the electromagnetic $N^*(1535) \to \gamma^* N$ transition. 
 Within the same formalism we evaluate the cross section for the reactions
 $\gamma N \to \eta N$. We find a fair agreement for the absolute values of the
 transition amplitudes, as well as
 for the $Q^2$ dependence of the amplitudes, within theoretical and experimental uncertainties discussed in the paper. The ratios obtained between the
 $S_{1/2}$  and $A_{1/2}$  for the neutron or proton states of the
 $N^*(1535)$ are in qualitative agreement with
 experiment and there is agreement on the signs. The same occurs for the
 ratio of cross sections for the $\eta $ photoproduction on neutron and
 proton targets in the vicinity of the $N^*(1535)$ energy. The global results
  support the idea of this resonance as being dynamically
 generated, hence, largely built up from meson baryon components. {However, the details of the model indicate that an admixture with a genuine quark state is also demanded that could help obtain a better agreement with experimental data.}

\end{abstract}

\maketitle

\section{Introduction}

The traditional picture of baryons as being made from three constituent quarks
\cite{quark} is giving rise, in some cases, to more complicated structures. One of
the ideas which has gained strength in recent times is that low lying
resonances of $J^P= 1/2^- , 3/2^-$ seem to be well represented in terms of
states which are generated by the meson baryon interaction in $L=0$; in the
$1/2^-$  case from the interaction of the octet of mesons of the $\pi$ with the
octet of baryons of the $p$
\cite{kaiser,angels,bennhold,joseulf,jido,nieves,carmen,hyodo} 
and in the $3/2^-$  from the interaction of the same
mesons with the decuplet of baryons of the $\Delta(1232)$ \cite{lutz,sarkar}. 
The $\Lambda(1405)$,
which actually comes as two poles in chiral theories 
\cite{jido}, with this two-pole structure 
supported by experiment \cite{magas}, has been for long thought of
as a kind of meson baryon molecule of the $\bar{K}N$ and $\pi \Sigma$ states 
\cite{Dalitz:1959dn,dalitz}, a structure similar to that provided by the chiral approaches
mentioned above.  The $N^*(1535)$ is one more resonance that appears in the two
octets and one singlet of dynamically generated resonances coming from the 
interaction of the octet of mesons of the $\pi$ with the octet of baryons of 
the $p$ \cite{jido}. In fact, it was noted earlier in \cite{siegel}, before the
systematics of \cite{jido} was established,  that the
interaction provided by chiral Lagrangians put as kernel of the Lippmann
Schwinger equation generated this resonance, which  also appears in other
work along similar lines \cite{arriola,inoue}. 

  The $N^*(1535)$ plays an important role
  in all processes of $\eta$ production since it couples very strongly to 
  $\eta N$. This feature is actually provided automatically by the chiral 
  theories, one of the points of support for the nature of this resonance as
  being dynamically generated. A recent study of the model dependence of the
  properties of this resonance is seen in \cite{dytman}.
  
    From the point of view of a dynamically generated resonance the $N^*(1535)$
 leads to fair descriptions of the $\pi N \to \eta N$ and $\gamma N \to \eta N$
 reactions \cite{siegel,inoue,michael} and produces reasonable numbers for the
 $\eta N$ scattering lengths \cite{kaiser,inoue}. Yet, it has been argued that
 one of the important tests of the nature of a resonance is its electromagnetic
 form factors. Indeed, a meson baryon resonance should get the $Q^2$ dependence
 basically from the meson cloud (we take as usual $Q^2=-k^2$ with $k$ being 
 the photon momentum). If this is a pion, this light particle has a
 fairly large extent in the wave function, as a consequence of which, the form
 factor of the resonance should fall relatively fast compared to ordinary quark
 models which confine the quarks at smaller distances. This is also the case for
 the proton at small $Q^2$, due to its meson cloud, which stabilizes later on at
 larger values of $Q^2$ where the quark components take over, as shown in chiral
 quark models \cite{thomas,tegen,amand}. We shall see that something special
 happens for the  $N^*(1535)$, but in any case this is a very stringent test,
 since the chiral theory provides the normalization and the $Q^2$ dependence for
 the different transition form factors without any free parameter, once the
 parameters used in $\pi N $ scattering with its coupled channels are fixed to
 scattering data. 
 
   Radiative decays of resonances from the point of view of their dynamically 
generated nature have been addressed in \cite{mishasourav} for the
$\Lambda(1520)$, in \cite{mishasolo} for the $\Delta(1700)$ and in
\cite{mishageng} for the two $\Lambda(1405)$ states. It concerns the decay of the
resonances into a baryon and a real photon.  Some work with virtual photons from
this point of view is
done in \cite{kaiser,Borasoy:2002mt} for the electroproduction of $\eta$  in the vicinity of
the $N^*(1535)$ resonance.  
Meanwhile, experimental analyses have succeeded in  extracting the helicity 
transition form factors for   $N^*(1535) \to N \gamma$ $A_{1/2}$ and
$S_{1/2}$, for both  $N=p,n$, in a relatively wide range of $Q^2$ values
\cite{burkert}.

  We evaluate these form factors from the point of view
of the  $N^*(1535)$ as a dynamically generated resonance. For that purpose we
shall extend the formalism of \cite{mishasolo,mishageng} to virtual photons. 
The new formalism requires changes from the real photon case, but it is
rewarding since it provides much more information, replacing the helicity
transition amplitudes by functions of $Q^2$ and adding the new $S_{1/2}$
transition form factor which only plays a role for virtual photons. Hence, there
is far more information to test the predictions of the model. 

  From the quark model point of view there has also been much work done on these
  helicity form factors \cite{close,Konen:1989jp,santopinto,metsch, capstick,capstickcont,Pace:1998pp,Warns:1989ie,Aiello:1998xq}.
  A comparison of their prediction with experiment plus a compilation of results
  from different experiments can be seen in
  \cite{burkert,Thompson:2000by,Aznauryan:2004jd}.  There are appreciable differences from one quark
  model to another and relativistic effects seem to be important, particularly
   in the $S_{1/2}$ helicity transition form factor. 
   {It should be noted that some of the models, particularly those incorporating relativistic effects \cite{metsch,capstick,Aiello:1998xq} produce a fair agreement with data, in particular  a good description of the $Q^2$ dependence of the form factor.}

   In our approach, the quarks enter through the meson and baryon components of
 the resonance and the $Q^2$ dependence is tied to the meson and baryon form
 factors, which we take from experiment, plus the particular $Q^2$ dependence of
 the loop functions from the meson baryon coupled channels that build up the 
resonance. Thus, the final $Q^2$ dependence is a non-trivial consequence of chiral
 dynamics, which provides the coupling of the resonance to open and closed
 channels, the $Q^2$ dependence of the different loops, and the form factors of
 the mesons and baryons, particularly the mesons, as we shall see. 
 
    The results that we obtain are in fair  agreement with experiment for both
charged states and for the two transition form factors, hence providing extra
support for the nature of the $N^*(1535)$ as being largely made from the
interaction of meson baryon coupled channels. 

{Some deficiency in the $Q^2$ dependence at large $Q^2$ could be an indication of a mixture of the meson baryon components with a genuine quark component, which is also indicated by particular details of the chiral approach that we shall mention below. }

\section{Helicity  amplitude}

\begin{figure}
\epsfxsize= 5cm
\begin{center}
  \epsfbox{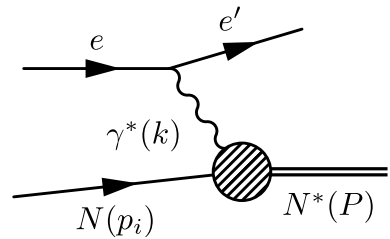}
\end{center}
  \caption{Kinematics of the electroproduction of $N(1535)$. \label{fig1}}
\end{figure}

We consider the production reaction of the $N(1535)$ resonance ($J^{p}=1/2^{-}$) by in-elastic electron-nucleon scattering as shown in Fig.\ref{fig1}. The $N(1535)$ is created by exchange of a virtual photon carrying momentum $k$. The initial $N$ and final $N^{*}$ momenta and masses are denoted by $(\pini,\Imass)$ and $(\pfin, \Fmass)$, respectively. The energy momentum conservation reads 
\begin{equation}
   \pfin = \pini + k . \label{eq:EMconv}
\end{equation}

There are two independent amplitudes for the electro-transition from 
$J^{P}=1/2^{+}$ to $1/2^{-}$, $A_{1/2}$ and $S_{1/2}$, which are defined in 
terms of the transition electric current $J_{\mu}$ by
\begin{eqnarray}
   A_{1/2} &=& \sqrt \frac{2\pi \alpha}{q_{R}} \frac{1}{e} 
   \langle N^{*}, J_{z}=\frac{1}{2} | \epsilon^{(+)}_{\mu} J^{\mu} | N, S_{z}=- \frac{1}{2} \rangle \\
   S_{1/2} &=& \sqrt \frac{2\pi \alpha}{q_{R}} \frac{1}{e}\frac{|\vec k|}{\sqrt{Q^{2}}} 
   \langle N^{*}, J_{z}=\frac{1}{2} | \epsilon^{(0)}_{\mu} J^{\mu} | N, S_{z}= \frac{1}{2} \rangle \ \ \
\end{eqnarray}
with the fine structure constant $\alpha = e^{2}/4\pi$, the energy equivalent {to that of} a real photon $q_{R} =(W^{2}-\Imass^{2})/(2W)$ and the photon-nucleon center-of-mass energy $W\equiv \sqrt{\pfin^{2}}$. The polarization vectors of the photon, $\epsilon_{\mu}$, are given by
\begin{eqnarray}
   \epsilon^{\pm}_{\mu} &=& \frac{1}{\sqrt 2} (0,\mp 1, -i ,0) \\
   \epsilon^{0}_{\mu} &=& \frac{1}{\sqrt {Q^{2}}} (k ,0, 0, -k^{0})    
\end{eqnarray}
with $Q^{2} = - k^{2}$, where we take the CM momenta $\vec k$ and $\vec \pini$ along the $z$ axis. 

Let us discuss the general expression of the transition current 
$J^{\mu}$ in the relativistic formulation. First of all, we recall the 
equation of motion for the initial nucleon 
\begin{eqnarray}
   (\pislash - \Imass) u_{i}(\pini) &=& 0  \label{eq:EOMforN}
\end{eqnarray}
where $u_{i}(\pini)$ is the Dirac spinor for the initial nucleon normalized by
\begin{equation}
   u_{N} =\sqrt{\frac{E_{i}+\Imass}{2\Imass}}\left(
     \begin{array}{c} 1 \\ \frac{\vec \sigma \cdot \vec \pini}{E_{i} + \Imass} \end{array}
     \right ) \chi .
\end{equation}
For the final $N^{*}$, we assume the pole dominance, so that we again have 
\begin{eqnarray}
   (\Pslash - \Fmass) u_{f} (P)  &=& 0 \label{eq:EOMforNs}
\end{eqnarray}
where $u_{f}(\pfin)$ is the $N^{*}$ Dirac spinor and $\Fmass$ denotes the real part of the $N^{*}$ mass. In the calculations of the helicity amplitude, the $\Fmass$ is chosen as the $N^{*}$ energy in the final state, $W \simeq 1535$ MeV. 

It follows that the terms involving the $\gamma$ matrix in $J_{\mu}$ are 
only {of the form} $\gamma \cdot \epsilon$, since we can move $\pislash$ and $\Pslash$ 
through $\gamma_{\mu}\gamma_{\nu}-\gamma_{\nu}\gamma_{\mu}=g_{\mu\nu}$ to 
either the left or right end in the amplitudes, and they can be replaced by 
the masses by means of Eqs.~(\ref{eq:EOMforN}) and (\ref{eq:EOMforNs}). The term 
involving $\kslash$ can  also be replaced by the momentum conservation $\kslash=\Pslash-\pislash$.
Thus,
Lorentz invariance and momentum conservation (\ref{eq:EMconv}) require the transition current $J^{\mu}$ to be written, in general, by the following three Lorentz scalar amplitudes: 
\begin{equation}
   J^{\mu} = (\Amp_{1} \gamma^{\mu} + \Amp_{2} P^{\mu} + \Amp_{3} k^{\mu})\gamma_{5} . \label{eq:Jmu}
\end{equation}

The gauge invariance $  k \cdot J = 0$, tells us that there are only two independent amplitudes among these three amplitudes, $\Amp_{i}$,  giving the following relation:
\begin{equation}
   (\Fmass+\Imass) \Amp_{1} + k\cdot P \Amp_{2} + k^{2} \Amp_{3} = 0  \label{eq:iden} \ .
\end{equation}
Using the transition current (\ref{eq:Jmu}), we evaluate the helicity amplitudes, $A_{1/2}$ and $S_{1/2}$, in the rest frame of the $N(1535)$ resonance. After some algebra,
the helicity amplitudes are written in terms of the amplitude $\Amp_{2}$ and $\Amp_{3}$ by 
\begin{eqnarray}
   A_{1/2} &=& \sqrt \frac{2\pi \alpha}{q_{R}} \sqrt{\frac{E_{i}+\Imass}{2\Imass}}
   \frac{1}{e}  \label{eq:AampR}
   \frac{ \sqrt{2} }{\Fmass+\Imass}
   \nonumber \\ && \ \ \ \ \times
    \left( k\cdot P \Amp_{2} + k^{2} \Amp_{3}\right)\\
   S_{1/2} &=&   \sqrt \frac{2\pi \alpha}{q_{R}} \sqrt{\frac{E_{i}+\Imass}{2\Imass}}\frac{1}{e} 
   \frac{-|\vec k| }{\Fmass+\Imass}
   \nonumber \\ && \ \ \ \ \times
   \left ( \Fmass\Amp_{2} + (\Fmass-\Imass) \Amp_{3}  \right)  \label{eq:SampR}
\end{eqnarray}

The transition current (\ref{eq:Jmu})  can be written equivalently in the CM frame as
\begin{eqnarray}
  J^{\mu} &=&\sqrt{\frac{E_{i}+M_{N}}{2M_{N}}}\left[ \Amp_{1} \sigma^{\mu} 
  \right. \nn \\ && \left.
  + \left( \frac{\Amp_{1}}{(E_{i}+M_{N})W} + \frac{\Amp_{2}}{E_{i}+M_{N}} \right) P^{\mu} \sigma \cdot k
  \right. \nn \\ && \left.
  + \frac{\Amp_{3}}{E_{i}+M_{N}} k^{\mu} \sigma \cdot k \right] \\ %\nonumber
  &\equiv& \Amp_{1}^{\rm NR} \sigma^{\mu} + \Amp^{\rm NR}_{2} P^{\mu} \sigma\cdot k + \Amp^{\rm NR}_{3} k^{\mu} \sigma \cdot k  \label{eq:nonreladecomp}
\end{eqnarray}
where $\sigma^{\mu} = (0, \vec \sigma)$ and we take the CM frame $P^{\mu}=(W , \vec 0)$. Then the helicity amplitudes are written in terms of the amplitudes defined above, $\Amp^{\rm NR}_{i}$, as
\begin{eqnarray}
   A_{1/2} &=& \sqrt \frac{2\pi \alpha}{q_{R}} 
   \frac{1}{e}
   \sqrt{2} \left( k\cdot P \Amp^{\rm NR}_{2} + k^{2} \Amp^{\rm NR}_{3}\right) \label{eq:AampNR}
   \\
   S_{1/2} &=&   \sqrt \frac{2\pi \alpha}{q_{R}}
\frac{-|\vec k| }{e} 
   \left (W \Amp^{\rm NR}_{2}+k^{0}\Amp^{\rm NR}_{3}    \right) \label{eq:SampNR}
\end{eqnarray}
with the gauge invariance condition 
for the nonrelativistic amplitudes
\begin{equation}
  \Amp_{1}^{\rm NR}  + \Amp^{\rm NR}_{2} k\cdot P + \Amp^{\rm NR}_{3} k^{2}  = 0. \label{eq:gauseinvNR}
\end{equation}

\section{Evaluation of the transition form factors}
\label{sec:formulation}
\subsection{Model of $N(1535)$ and photon coupling}
\label{sec:model}
In our approach, the $N(1535)$ resonance is dynamically generated in the $s$-wave meson baryon scattering in the coupled channels of $\pi^{-} p$, $\pi^{0} n$, $\eta n$, $K^{+} \Sigma^{-}$, $K^{0} \Sigma^{0}$, $K^{0} \Lambda$ for the neutron resonance (with neutral charge) and $\pi^{0} p$, $\pi^{+} n$, $\eta p$, $K^{+} \Sigma^{0}$, $K^{0} \Sigma^{+}$, $K^{+} \Lambda$ for the proton resonance (with $+1$ charge). 
The scattering amplitude for the $N(1535)$ resonance is described in Ref.\cite{inoue} by means of the Bethe-Salpeter equation for meson baryon scattering  given by
\begin{equation}
   T = V + V G T \ .
\end{equation}
Based on the $N/D$ method and the dispersion relation \cite{joseulf}, this integral scattering equation can be reduced to a simple algebraic equation
\begin{equation}
   T = (1-VG)^{-1}\,V
   \label{bse}
\end{equation}
where the matrix $V$ is the $s$-wave meson-baryon interaction given by the lowest order of the chiral perturbation theory, which is the Weinberg-Tomozawa interaction, 
given by
\begin{eqnarray}
V_{i j} &=&
 - C_{i j} \frac{1}{4 f^2}(2\sqrt{s} - M_{i}-M_{j})
 \nonumber \\ &&  \ \ \ \times
\sqrt{\frac{M_{i}+E}{2M_{i}}}
\sqrt{\frac{M_{j}+E^{\prime}}{2M_{j}}}
\label{eq:ampl2}
\end{eqnarray}
with the channel indices $i,j$, the baryon mass $M$, the meson mass $m$, the meson decay constant $f$  and the center of mass energy $\sqrt s$. The coefficient $C_{ij}$ is the coupling strength of the meson and baryon, which is determined by the SU(3) group structure of the channel. 
The diagonal matrix $G$ is the meson baryon loop function given in terms of the meson and baryon propagators by
\begin{eqnarray}
   G(\sqrt s) &=&  i\int \frac{d^{4} q}{(2\pi)^{4}} \frac{M}{E(\vec q)}
    \frac{1}{q^{0} -E(\vec q) + i\epsilon}
   \nonumber \\ && \ \ \ \ \ \ \times
    \frac{1}{(P-q)^{2} - m^{2} + i \epsilon}
    \label{propnorela}
\end{eqnarray}
with  the total energy $P=(\sqrt s, 0, 0,0)$ in the center of mass frame. For the baryon propagator we use the nonrelativistic form and  neglect the negative energy propagation. The loop function should be regularized with proper schemes. In the practical calculation, we take dimensional regularization by using a covariant form of the positive energy part of the baryon propagator,
\begin{eqnarray}
\frac{M}{E(\vec q\, )}\,\frac{\Sigma_r u_r(\vec q\, )\overline{u}_r(\vec q\, )}{q^0-E(\vec q\,)+i\epsilon}\simeq
\frac{2M \,\Sigma_r u_r(\vec q\, )\overline{u}_r(\vec q\, )}{q^2-M^2+i\epsilon}.
\label{covaform}
\end{eqnarray}
In dimensional regularization,  the loop function in each channel $i$ is given by the following analytic expression:   
\begin{eqnarray}
G_i &=& i \,  \int \frac{d^4 q}{(2 \pi)^4} \,
\frac{2 M_i}{q^2 - M_i^2 + i \epsilon} \, \frac{1}{(P-q)^2 - m_i^2 + i
\epsilon}  \nonumber \\ &=& \frac{2 M_i}{16 \pi^2} \left\{ a_i(\mu) + \ln
\frac{M_i^2}{\mu^2} + \frac{m_i^2-M_i^2 + s}{2s} \ln \frac{m_i^2}{M_i^2} +
\right. \nonumber \\ & &  \phantom{\frac{2 M_i}{16 \pi^2}} +
\frac{\bar q_i}{\sqrt{s}}
\left[
\ln(s-(M_i^2-m_i^2)+2 \bar q_i\sqrt{s})
\right. \nonumber  \\
&&  \phantom{\frac{2 M_i}{16 \pi^2} +\frac{\bar q_i}{\sqrt{s}}}
  \hspace*{-0.6cm}+ \ln(s+(M_i^2-m_i^2)+2 \bar q_i\sqrt{s}) \nonumber  \\
& & \phantom{\frac{2 M_i}{16 \pi^2} +\frac{\bar q_i}{\sqrt{s}}}
  \hspace*{-0.6cm}- \ln(-s+(M_i^2-m_i^2)+2\bar q_i\sqrt{s}) \nonumber  \\
& & \left. \left. 
     \phantom{\frac{2 M_i}{16 \pi^2} +\frac{\bar q_i}{\sqrt{s}}}
 \hspace*{-0.6cm} - \ln(-s-(M_i^2-m_i^2)+2\bar q_i\sqrt{s}) \right]
\right\} ,
\label{eq:gpropdr}
\end{eqnarray}
where $\bar q_i$ is the 3-momentum of the meson or baryon in the center of mass frame, $\mu$ is the scale of dimensional regularization and $a_i(\mu)$ are subtraction constants, which are determined by a fit to the $S_{11}$ and $S_{31}$ partial waves of $\pi N$ scattering \cite{inoue}. {Once these constants are fixed to the $\pi N$ scattering data, the amplitudes involving photons can be predicted without introducing any new free parameters.}

{It should be emphasized that the subtraction constants $a_i(\mu)$ are different for different channels $i$ in the model of \cite{inoue}. This is unlike the case of $\overline{K}N$ scattering and the $\Lambda(1405)$ resonance, where all the subtraction constants in the different channels are approximately equal and of natural size according to Ref. \cite{joseulf}. The need for different subtraction constants in the case of $\pi N$ scattering and the $N^*(1535)$ resonance has been interpreted recently \cite{Hyodo:2008xr} as a clear indication that the $N^*(1535)$ contains a mixture of a genuine quark state apart from the meson baryon components. This conclusion has been reached by following an alternative method in which the subtraction constants have been chosen of natural order, and approximately equal, and a CDD pole is included which would give us an indication that extra components to the meson baryon ones are needed in the $N^*(1535)$ wave function. The study of Ref. \cite{Hyodo:2008xr} clearly indicates that the effect of the CDD pole is negligible for the $\Lambda(1405)$ resonance but relevant for the the case of the $N^*(1535)$. We shall see that our approach, based on the meson baryon components exclusively, provides a fair description of data, but some remaining discrepancies indirectly hint to the need of extra components in the wave function.}

The resulting amplitudes $T^{ij}$ from Eq.~(\ref{bse}) can be analytically continued to the complex plane of the scattering energy $s^{1/2}$. The amplitude has a pole in the complex plane that is identified with the resonance, and the coupling strengths $g_{i}$ of the resonance to the meson-baryon channels is determined by the residues of the pole:
\begin{equation}
   T_{N^{*}}^{ij}(\sqrt{s}) = \frac{g_{i}g_{j}}{\sqrt{s} - M_{R} + i\Gamma_{R}/2} + T_{\rm BG}^{ij}
\end{equation}
where $\sqrt{s}$ is the c.m. energy of the meson-baryon system and $T_{\rm BG}^{ij}$ is an amplitude for the nonresonant contributions. 
The pole positions of the resonance are obtained as 
\begin{equation}
   \sqrt s = 1537 -37i  \ \ [\MeV]  \label{eq:PPnstar}
\end{equation}
for the $n^{*}$ (neutral charge) and 
\begin{eqnarray}
   \sqrt s = 1532 -37i \ \ [\MeV] \label{eq:PPpstar}
\end{eqnarray}
for the $p^{*}$ ($+1$ charge).
The values of the coupling constants are listed in Table \ref{tab:coupling}. The coupling constants $g_{i}$ characterize the structure of the $N^{*}$. The empirical evidence of larger coupling of the $N(1535)$ to $\eta N$ than that to $\pi N$ is reproduced in this model. In addition, the couplings to the $\Sigma K$ and $\Lambda K$ channels are also large. This implies that the $N(1535)$ has large components of strangeness.

\begin{table}
\caption{Complex coupling constants {$g_i$} of $n^{*}$ to the meson-baryon channels. \label{tab:coupling}}
\begin{center}
\begin{tabular}{ccc}
 \hline\hline
 $\pi^{-} p$& $\pi^{0} n$& $\eta n$ \\
 $ 0.557+ 0.325 i $ & $-0.387- 0.238 i $ & $ -1.45+  0.435 i$ \\ 
 \hline
  $K^{+} \Sigma^{-}$& $K^{0} \Sigma^{0}$& $K^{0} \Lambda$ \\
 $ 2.20  - 0.171 i$& $-1.56+  0.115 i$ & $ 1.39  - 0.0825 i$ \\
  \hline\hline
\end{tabular}
\end{center}
\label{tab:ncoupl}
\end{table}

\begin{table}
\caption{Coupling constants {$g_i$} of $p^{*}$ to the meson-baryon channels. \label{tab:coupling2}}
\begin{center}
\begin{tabular}{ccc}
 \hline\hline
 $\pi^{0} p$& $\pi^{+} n$& $\eta p$ \\
  $ 0.397+ 0.222 i$&$ 0.555+ 0.322 i$&$-1.47+ 0.432 i$\\
 \hline
  $K^{+} \Sigma^{0}$& $K^{0} \Sigma^{+}$& $K^{+} \Lambda$ \\
  $ 1.56 - 0.133 i$& $ 2.21 - 0.183 i$&$ 1.37 - 0.100 i$ \\
  \hline\hline
\end{tabular}
\end{center}
\label{tab:pcoupl}
\end{table}

\begin{figure}
\epsfxsize= 8.5cm
\begin{center}
  \epsfbox{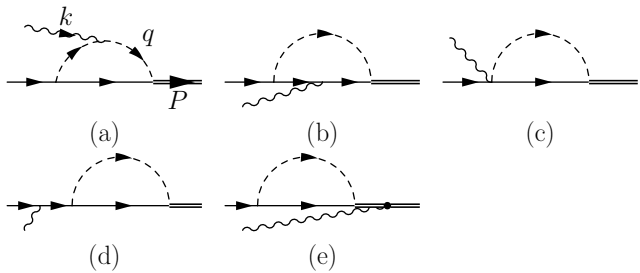}
\end{center}
  \caption{Feynman diagrams for the transition form factor of $N(1535)$ at one loop level. The solid, dashed, wavy and double lines denote octet baryons, mesons, photon and $N(1535)$, respectively. \label{fig:FeynmanDiagram}}
  \label{fig:loops}
\end{figure}

In the meson-baryon picture of the $N(1535)$ resonance, the photoproduction of the resonance from the nucleon is formulated through the photon couplings to the meson and baryon {components of the $N^*(1535)$}. 
Photon couplings and gauge invariance in the case of chiral unitary amplitudes are discussed in Ref.\cite{Nacher:1999ni,Borasoy:2005zg,Borasoy:2007ku}. Here, we follow 
an approach similar to the one developed in Refs.\cite{mishasourav, mishasolo, mishageng} for real photons, extending it to virtual ones. 
%%%%%%
%
Feynman diagrams to the transition form factors at one-loop level are shown in Fig.~\ref{fig:FeynmanDiagram}. In the loops, all possible octet mesons and baryons contribute, namely,  $\pi^{-} p$, $\pi^{0} n$, $\eta n$, $K^{+} \Sigma^{-}$, $K^{0} \Sigma^{0}$, $K^{0} \Lambda$ for the neutron resonance (neutral charge) and $\pi^{0} p$, $\pi^{+} n$, $\eta p$, $K^{+} \Sigma^{0}$, $K^{0} \Sigma^{+}$, $K^{+} \Lambda$ for the proton resonance ($+1$ charge).
 We sum up all the contributions to the transition amplitudes.  In diagrams
  (a) and (b), the photon attaches to the meson and baryon in the loop, 
  respectively. Diagram (c) has the Kroll-Ruderman coupling which is the contact interaction of the photon, meson and baryon. 
 Diagrams (d) and (e) have to be taken into account to keep gauge invariance. 
 It seems that these diagrams contain a forbidden transition from the 
 $1/2^{+}$ state of the proton to the  $1/2^{-}$ state of the $N(1535)$, but  
 negative energy propagation of the intermediate baryons in these diagrams 
 is possible, having the opposite parity to the positive energy propagation. 
 The positive energy part in motion also mixes different parity states through the different partial waves. 
 
We calculate the transition amplitudes both in a non relativistic and relativistic formulations. The momentum of the baryon is small enough to describe the transition amplitudes in nonrelativistic formulation. In addition, as we have already mentioned above, in the construction of the $N(1535)$ in the meson-baryon scattering \cite{inoue}, we have used the nonrelativistic formulation in the elementary vertex and the baryon propagators as seen in Eqs.~(\ref{eq:ampl2}) and (\ref{eq:gpropdr}). Therefore, to keep consistency of the calculation of the photon couplings with the construction of the $N^{*}$ resonance, the nonrelativistic calculation is preferable. 
   One should mention that the fact, that one fits subtraction constants to
   data, largely washes out relativistic effects from the use of Eq.\eqref{propnorela} or
   \eqref{eq:gpropdr} in the $G$ function in a fair range of energies around the fitted point.
Nevertheless, it is somewhat complicated to prove gauge invariance in the nonrelativistic formalism, since we need to take 
into account all the possible diagrams including negative energy contributions, which are referred to as Z-diagrams. To avoid this 
complication, we will also perform the calculation of the amplitudes in relativistic formulation, in which the negative energy contributions are automatically counted without introducing the Z-diagrams, and we shall show that the relativistic calculation is exactly gauge invariant. 
This guarantees  that each term in the $1/M$ expansion is gauge invariant.
Exploiting this fact, in the nonrelativistic framework we calculate diagrams for leading amplitudes relying upon gauge invariance and show that 
the next-to-leading terms are relatively  small.

The basic interactions of the mesons and baryons are given by the following chiral Lagrangian: 
\begin{eqnarray}
{\cal L}_{MBB} &=& 
 - \frac{D}{\sqrt 2 f} \, {\rm Tr} \left[ \bar B \gamma_{\mu} \gamma_{5} \{\partial^{\mu}\Phi,B\} \right]
 \nonumber \\ && 
 - \frac{F}{\sqrt 2 f}\, {\rm Tr} \left[ \bar B \gamma_{\mu} \gamma_{5} [\partial^{\mu}\Phi ,B] \right] \label{eq:MBcoup}
\end{eqnarray}
with the meson and baryon fields, $\Phi$ and $B$, defined by
\begin{eqnarray}
   \Phi &=& 
   % {1 \over \sqrt{2}} \lambda^a \Phi^a =
    \left(
   \begin{array}{ccc}
       \frac{1}{\sqrt{2}} \pi^0 + \frac{1}{\sqrt{6}} \eta & \pi^+ & K^+ \\
       \pi^- & -\frac{1}{\sqrt{2}} \pi^0 + \frac{1}{\sqrt{6}} \eta  & K^0\\
       K^- & \bar K^0 & - \frac{2}{\sqrt{6}} \eta
   \end{array}
   \right)
\\
   B &=& 
 %  {1 \over \sqrt{2}} \lambda^a B^a = 
   \left(
   \begin{array}{ccc}
       \frac{1}{\sqrt{2}} \Sigma^0 + \frac{1}{\sqrt{6}} \Lambda &
       \Sigma^+ & p \\
       \Sigma^- & -\frac{1}{\sqrt{2}} \Sigma^0 + 
       \frac{1}{\sqrt{6}} \Lambda  & n\\
       \Xi^- & \Xi^0 &- \frac{2}{\sqrt{6}} \Lambda
   \end{array}
   \right) \ .
\end{eqnarray}
%0829
The $MBB$ couplings from these Lagrangian are given by $g_{A}^{i}/(2f)$ with the corresponding axial vector coupling $g_{A}^{i}$ and the meson decay constant $f$. The axial vector couplings are given in terms of the $D$ and $F$ parameters in the Lagrangian (\ref{eq:MBcoup}) as listed in Table \ref{tab:gA}. 
For the meson decay constant, we use $f=93$ MeV for all channels in our calculation. The values of $D$ and $F$ for the axial vector couplings are taken from Ref.\cite{Luty:1993gi} as
\begin{equation}
D=0.85 \pm 0.06 \ ,\ \ \ \ \ F=0.52 \pm 0.04 \ .  \label{eq:DFvalues}
\end{equation}
These values are determined by the experimental data of the hyperon axial vector couplings, neglecting higher-order corrections. 
The photon couplings to  mesons and  baryons are given by the gauge couplings:
\begin{eqnarray}
    {\cal L}_{\gamma B} &=& -e {\rm Tr}\left[ \bar B \gamma_{\mu} [Q_{\rm ch},B]\right] A^{\mu} \label{eq:LaggBB} \\
    {\cal L}_{\gamma M} &=& ie {\rm Tr}\left[ \partial_{\mu} \Phi [Q_{\rm ch},\Phi]\right] A^{\mu}
\end{eqnarray}
with the charge matrix $Q_{\rm ch}={\rm diag}(\frac{2}{3},-\frac{1}{3},-\frac{1}{3})$.
The Kroll-Ruderman terms, the $\gamma MBB$ couplings, are obtained by  replacing the derivative acting on the meson fields, $\partial_{\mu}\Phi$, with the covariant derivative $D_{\mu}\Phi = \partial_{\mu}\Phi + i e A_{\mu} [Q_{\rm ch},\Phi]$ in the interaction Lagrangian (\ref{eq:MBcoup}) to realize the gauge invariance. The Kroll-Ruderman terms are proportional to the meson charge $Q_{M}$. For the couplings of the meson-baryon to the $N^{*}$ resonance, we take a Lorentz scalar form representing the $s$-wave nature and the coupling strengths are taken from the chiral unitary approach as given in Tables \ref{tab:coupling}, \ref{tab:coupling2}.

\begin{table}
\caption{The axial vector coupling $g_{A}^{i}$ for each channel. The values of $D$ and $F$ are given in Eq.~(\ref{eq:DFvalues}).  \label{tab:gA}
}
\begin{tabular}{cccc}
\hline
 channel &  
$np\pi^{-}$ & $nn\pi^{0}$ & $nn\eta$  \\
 $g_{A}^{i}$ & 
 $\sqrt 2 (D+F)$ & $- D-F$ & $\frac{1}{\sqrt 3} (-D + 3 F)$\\
\hline
 & $n\Sigma^{-}K^{+}$ &$n\Sigma^{0} K^{0}$& $n\Lambda K^{0}$ \\
 & $\sqrt 2( D -  F)$ & $- D+F$ & $- \frac{1}{\sqrt 3} (D + 3 F)$ \\
 \hline\hline
 channel &  
$pp\pi^{0}$ & $pn\pi^{-}$ & $pp\eta$  \\
 $g_{A}^{i}$ &  
 $D+F$ & $\sqrt{2}( D+F) $ & $\frac{1}{\sqrt 3} (-D + 3 F)$ \\
 \hline
&$p\Sigma^{0}K^{+}$ & $p\Sigma^{+} K^{0}$& $p\Lambda K^{+}$ \\
& $ D -  F$ & $\sqrt 2(D-F)$ & $- \frac{1}{\sqrt 3} (D + 3 F)$  \\
\hline
\end{tabular}
\end{table}

\subsection{Nonrelativistic formulation}
\label{sec:norela}

\begin{figure}[t]
\epsfxsize= 8.5cm
\begin{center}
  \epsfbox{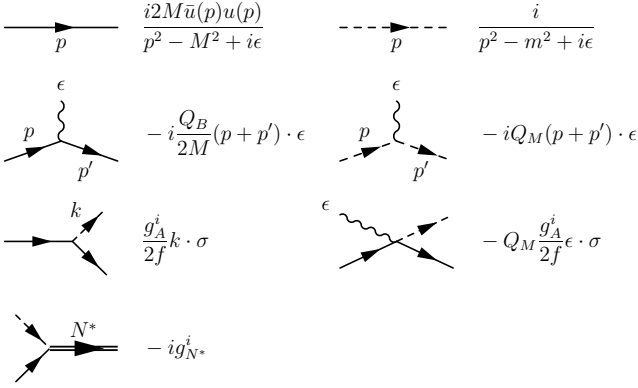}
\end{center}
  \caption{Nonrelativistic Feynman rules for the propagator and the elementary vertices. The solid, dashed, wavy and double lines denote octet baryons, mesons, photon and $N(1535)$, respectively. $M$ and $m$ denote the baryon and meson masses, respectively. $Q_{B}$ and $Q_{M}$ are the charges of the baryon and meson. \label{fig:FeynmanRule}}
\end{figure}

In the  nonrelativistic formulation, the leading terms of the $1/M$ expansions 
are the 
diagrams (a) and (c) in Fig.\ref{fig:FeynmanDiagram}. Diagram (b) is found to 
be of next-to-leading order due to the $1/M$ factor in the $\gamma BB$ coupling.  
In the CM frame of the $N^{*}$, which we take for the nonrelativistic 
calculation, diagram (d) vanishes, since there is a direct transition 
of $1/2^{+}$ to $1/2^{-}$. Diagram (e) has some contribution in this frame, but it is also  found to be of next-to-leading order, since the contribution to diagram (e) is confirmed to vanish in the large $M$ limit.
Indeed, if one neglects the kinetic energy term of the baryon propagators in 
the loop $(\vec p^{\, 2} / 2 M)$, the loop function vanishes. It also vanishes 
in the rest frame of the nucleon since it again involves a $1/2^{+}$ to $1/2^{-}$ transition. 

We will obtain each component of the diagram (a) and (c) in the decomposition in terms of the Lorentz structure given in Eq.~(\ref{eq:nonreladecomp}). Since the helicity amplitudes can be expressed by the $\Amp_{2}^{\rm NR}$ and $\Amp_{3}^{\rm NR}$ in  Eqs.~(\ref{eq:AampNR}) and (\ref{eq:SampNR}), we will calculate only these two amplitudes. The amplitudes $\Amp_{2}^{\rm NR}$ and $\Amp_{3}^{\rm NR}$ remain finite even with one loop integration.  In fact, $\Amp_{1}^{\rm NR}$ does have divergence in the loop calculation, which should cancel with divergences coming from the other diagrams thanks to gauge invariance. Here we do not confirm the cancellation of the divergences, since we later show the complete cancellation in the relativistic formulation.

The Feynman rules for the nonrelativistic couplings are summarized in Fig.~\ref{fig:FeynmanRule}. In the figure, $\epsilon^{\mu}$ denotes the photon polarization, and $\sigma$ is a Lorentz covariant form of the spin matrix, $\sigma^{\mu} = (0,\vec \sigma)$. $g_{A}^{i}$ stands for the axial vector coupling constants of the baryons to the corresponding meson. The values of $g_{A}^{i}$ for each channel are given in Table \ref{tab:gA}. 
$g_{N^{*}}^{i}$ is the coupling strength of the $N^{*}$ to the meson-baryon channel $i$. The values of $g_{N^{*}}^{i}$ are listed in Table \ref{tab:ncoupl} and \ref{tab:pcoupl}.
For the baryon propagator we use the covariant form  from Eq. (\ref{covaform}). 
The $\gamma BB$ coupling is used for the calculation of diagram (b), which is not taken into account in our final result of the nonrelativistic calculation. But in order to confirm that sub-leading terms from the $1/M$ expansion are negligibly small, we have calculated the diagram (b) in the nonrelativistic calculation. The $\gamma BB$ vertex is obtained by a nonrelativistic reduction of the interaction Lagrangian (\ref{eq:LaggBB}) as
\begin{eqnarray}
   -i Q_{B} \bar u \gamma \cdot \epsilon u &\rightarrow& - iQ_{B} \chi^{\dagger} \left[ \epsilon^{0} - \frac{\vec\epsilon \cdot (\vec p + \vec p^{\, \prime} )}{2M} \right] \chi   \label{eq:NRgBB1} \\
   &\simeq&  - iQ_{B} \chi^{\dagger} \left[  \frac{\epsilon \cdot (p + p^{\, \prime} )}{2M} \right ] \chi
   \label{eq:NRgBB2}
\end{eqnarray}
where we have used the fact that the baryon kinetic energies are small in the nonrelativistic kinematics, $p^{0} \simeq p^{0\prime} \simeq M$,  in the last expression. 
In Eqs.~(\ref{eq:NRgBB1}) and (\ref{eq:NRgBB2}) we have neglected the magnetic term that behaves like $(\vec \sigma \times \vec k) / 2M $ which has one power less in the loop variable. 
In Sec. \ref{sec:hiorder}, we shall  estimate the contributions from the convection current and the magnetic terms. 
In Eqs.~(\ref{eq:NRgBB1}) and (\ref{eq:NRgBB2}), 
$Q_{B}$ is the baryon charge such that it is $e$ for the proton
with $e^{2}/(4\pi) = \alpha \simeq 1/137$.

Let us start with  diagram (a). Applying the Feynman rules shown in Fig.\ref{fig:FeynmanRule}, the amplitude 
($-it=J\cdot \epsilon$)
for channel $i$ is calculated as
\begin{widetext}
\begin{eqnarray}
   -it_{a}^{i} &=& \int \frac{d^{4} q}{(2\pi)^{4}} 
   (-ig_{N^{*}}^{i})
   \frac{i2M_{i}}{(P-q)^{2} - M^{2}_{i}+i\epsilon} 
   \left( \frac{g_{A}^{i}}{2f} \right) (q - k) \cdot \sigma 
   \frac{i}{q^{2} - m^{2}_{i} + i\epsilon} 
   (-i Q_{M}) (2q-k)\cdot \epsilon
   \frac{i}{(q-k)^{2} - m^{2}_{i} + i\epsilon} 
   \nonumber \\
   & = & 
   iQ_{M} A_{i}   \int \frac{d^{4} q}{(2\pi)^{4}} 
   \frac{  ( q - k) \cdot \sigma  \,   (2q-k)\cdot \epsilon}
   {((P-q)^{2} - M^{2}_{i}+ i\epsilon)(q^{2} - m^{2}_{i} + i\epsilon)((q-k)^{2} - m^{2}_{i}+ i\epsilon )} 
\end{eqnarray}
\end{widetext}
where the coefficient $A_{i}$ is defined by
\begin{equation}
 A_{i}= \frac{  g_{A}^{i} g_{N^{*}}^{i} M_{i}}{f }\ . \label{eq:defA}
\end{equation}
We use the Feynman parameterization of the integral
\begin{equation}
   \frac{1}{abc} = 2 \int^{1}_{0} dx \int^{x}_{0} dy \frac{1}{\left(a + (b-a) x + (c-b)y\right)^{3}} \ . \label{eq:FeynParaInt}
\end{equation}
Then, using the integral variable $q^{\prime}$, such that $q=q^{\prime}+P(1-x) + ky$ and renaming $q^{\prime}$ as $q$, we eliminate the linear terms of $q$  in the denominator and obtain
\begin{widetext}
\begin{eqnarray}
   -it_{a}^{i}   & = & 
 iQ_{M} A_{i} 2 \int^{1}_{0} dx \int^{x}_{0} dy 
%   \nonumber \\ && \ \ \ \ \ \ \ \
   \int \frac{d^{4} q}{(2\pi)^{4}} 
   \frac{  ( q +(y-1) k) \cdot \sigma  \,   (2q+(2y-1)k +2(1-x) P)\cdot \epsilon}
   { \left(q^{2} - S_{a}^{i} + i\epsilon \right)^{3}} \ ,  \label{eq:diaaNR}
\end{eqnarray}
\end{widetext}
where we use $P \cdot \sigma = 0$ in the CM frame and $S_{a}^{i}$ is defined by
\begin{eqnarray}
   S_{a}^{i} &=&  2 P\cdot k (1-x)y - P^{2}x(1-x) -  k^{2} y (1-y)   \nn \\ &&
   + M^{2}_{i} (1-x) + m_{i}^{2}x \ . \label{eq:defSa}
\end{eqnarray}
In Eq.~(\ref{eq:diaaNR}), even powers of $q$ give contributions after performing the integration. The $q^{\mu}q^{\nu}$ term in the numerator which contributes to the $\Amp^{\rm NR}_{1}$ is divergent, while the terms with 0th power of $q$ remain finite and contribute to the $\Amp^{\rm NR}_{2}$ and $\Amp^{\rm NR}_{3}$ amplitudes. Finally after the integration, we obtain the $\Amp^{\rm NR}_{2}$ and $\Amp^{\rm NR}_{3}$ components  for the channel $i$ as
\begin{eqnarray}
    \Amp^{i \rm (NR)}_{2a} &=&    \frac{ Q_{M} A_{i}}{(4\pi)^{2}} 
    \int^{1}_{0} dx \int^{x}_{0} dy  \frac{2(y-1)(1-x)}{S_{a}^{i} - i\epsilon } \label{eq:amp2NR}
    \\
    \Amp^{i \rm (NR)}_{3a} &=&   \frac{ Q_{M} A_{i}}{(4\pi)^{2}}
    \int^{1}_{0} dx \int^{x}_{0} dy  \frac{(y-1)(2y-1)}{S_{a}^{i} - i\epsilon } \label{eq:amp3NR}
\end{eqnarray}
where we have used
\begin{eqnarray}
  \intq \frac{1}{(q^{2}-S)^{3}} &=& -\frac{i}{(4\pi)^{2}} \frac{1}{2}
  \left(\frac{1}{S}\right)\ .  \label{eq:LoopS3}
\end{eqnarray}

In a similar way we evaluate the contribution from diagram (b) which, 
as we mentioned, is of order $1/M$ of the previous ones and we obtain
\begin{eqnarray}
    \Amp^{i \rm (NR)}_{2b} &=&   -\frac{ Q_{B} A_{i}}{(4\pi)^{2}} 
    \int^{1}_{0} dx \int^{x}_{0} dy  \frac{2y(1-x)}{S_{b}^{i}-i\epsilon}\non
    \Amp^{i \rm (NR)}_{3b} &=&   - \frac{ Q_{B} A_{i}}{(4\pi)^{2}}
    \int^{1}_{0} dx \int^{x}_{0} dy  \frac{y(2y-1)}{S_{b}^{i}-i \epsilon}
    \label{convection}
\end{eqnarray}
with
\begin{eqnarray}
   S_{b}^{i} &= & 2 P\cdot k (1-x)y - P^{2}x(1-x) -  k^{2} y (1-y) 
   \nn \\ && 
   + m^{2}_{i} (1-x) + M_{i}^{2}x \ . \label{eq:defSb}
\end{eqnarray}

For the diagram (c), the amplitude has only the $\Amp^{\rm NR}_{1}$ component as seen in
\begin{widetext}
\begin{eqnarray}
   -it_{c}^{i} &=& (-ig_{N^{*}}^{i}) \int \frac{d^{4} q}{(2\pi)^{4}} 
   \frac{i 2M}{(P-q)^{2} - M^{2} + i\epsilon} 
   ( -Q_{M} \frac{g_{A}^{i}}{2f} )  \epsilon \cdot \sigma 
    \frac{i}{q^{2} - m^{2}+i\epsilon} \nonumber \\
   &=&    
  i Q_{M} A_{i}
   \int \frac{d^{4} q}{(2\pi)^{4}} 
   \frac{ \epsilon \cdot \sigma }
   {((P-q)^{2} - M^{2} + i\epsilon)(q^{2} - m^{2}+i\epsilon)}\ .
\end{eqnarray}
\end{widetext}
Hence we do not need to perform further calculation for this amplitude. 

Finally the helicity amplitudes in the nonrelativistic formulation are obtained by summing up all the channels and substituting the amplitudes (\ref{eq:amp2NR}) and (\ref{eq:amp3NR}) in Eqs.~(\ref{eq:AampNR}) and (\ref{eq:SampNR}).

\subsection{Relativistic formulation}
\label{sec:relacal}

\begin{figure}[t]
\epsfxsize= 8.5cm
\begin{center}
  \epsfbox{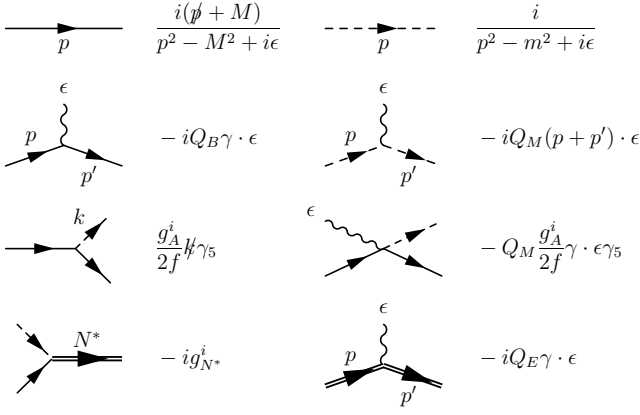}
\end{center}
  \caption{
  Same as Fig.\ref{fig:FeynmanRule} for the relativistic formulation. $Q_{E}$ stands for the external $N^{*}$ charge. 
  \label{fig:FeynmanRuleRela}}
\end{figure}

In this subsection, we calculate the transition amplitudes in relativistic 
formulation at one loop level. One of our purposes for the relativistic 
calculation is to confirm gauge invariance of our formulation. Without 
the $1/M$ expansion, which has been performed in the nonrelativistic 
calculation, all the diagrams shown in Fig.\ref{fig:FeynmanDiagram} should
 be calculated to make the amplitudes gauge-invariant at one loop level. 
 Each diagram has divergence from the loop integral. It will be found that the divergence appears only in the $\Amp_{1}$ term. 
% again\com{What means here "again"?}.  
After testing gauge invariance by  summing up all the diagrams, the amplitudes should be  finite without any regularization, since gauge invariance assures cancellation of the divergences coming from each diagram.  
In order to check this cancellation of the divergences, we calculate the 
$\Amp_{1}$ terms from all the diagrams.
 To isolate the divergent parts of the amplitudes, we exploit dimensional regularization, which respects gauge-invariance. The calculations are done in $d$-dimension, and then we expand $d$ around $d=4$ in terms of $\varepsilon$ given by $d=4-2\varepsilon$.  We also calculate the finite $\Amp_{2}$ and $\Amp_{3}$ to obtain the helicity amplitudes in the relativistic formulation. 
We will find again that only  diagrams (a) and (b) contribute to the 
$\Amp_{1}$ and $\Amp_{2}$ amplitudes.

Let us start with diagram (a). Using the Feynman rules shown in Fig.\ref{fig:FeynmanRuleRela}, the amplitude of the diagram (a) for the channel $i$ is given by
\begin{widetext}
\begin{eqnarray}
   -iT_{a}^{i} &=&     \intqd
   (-ig_{i}) 
 \prop{\Pslash-\qslash + \Bmass}{(P-q)^{2} - \Bmass^{2}}
 \left( \frac{g_{A}^{i}}{2\fpi} \right)(\qslash - \kslash) \gamma_{5} 
 \prop{1}{(q-k)^{2}-\Mmass^{2}}   \prop{1}{q^{2}-\Mmass^{2}} 
  (-iQ_{M}) (2 q -k) \cdot \epsilon \nn \\
 &=& i Q_{M} B_{i}  \intqd
 \frac{(\Pslash-\qslash + \Bmass)(\qslash - \kslash) \gamma_{5}  (2 q -k) \cdot \epsilon}
 {((P-q)^{2} - \Bmass^{2})((q-k)^{2}-\Mmass^{2}) (q^{2}-\Mmass^{2})} \label{eq:Ta}
\end{eqnarray}
\end{widetext}
where we define $B_{i}= (g_{A}^{i} g_{N^{*}}^{i}/2f)$. 
We can write expressions for all the other diagrams and then an explicit calculation shows that by substituting $\epsilon^{\mu}$ by $k^{\mu}$ one obtains an exact cancellation of the terms, hence passing the ordinary test of gauge invariance. The coupling of the photon to all lines and vertices in the meson baryon loop diagrams guarantees gauge invariance, as is commonly known \cite{Borasoy:2005zg,JuliaDiaz:2006xt}. 

We now come back to the amplitude $T_{a}$ of Eq.~(\ref{eq:Ta}) and, 
using the Feynman parameter integral (\ref{eq:FeynParaInt}), shifting the integral variable $q$ to $q=q^{\prime} + P(1-x) +ky$ and renaming $q^{\prime}$ by $q$, we find the integrand written as a function of $q^{2}$ as
\begin{eqnarray}
   -iT_{a}^{i}   & = & 
 iQ_{M} B_{i} 2 \int^{1}_{0} dx \int^{x}_{0} dy 
   \intqd
   \frac{ \textfrac{1}{2}  q^{2} C_{a} + D_{a}}
   { \left(q^{2} - S_{a}^{i} + i\epsilon \right)^{3}} \nonumber \ \\  \label{eq:diaaR}
\end{eqnarray}
where $S_{a}^{i}$ is defined in Eq.~(\ref{eq:defSa}) and the coefficients $C_{a}$ and $D_{a}$ are defined   in terms of the Lorentz components by 
\begin{eqnarray}
   C_{a} &=&  C_{a1} \eslash\gamma_{5}  + C_{a2}  P \cdot \epsilon \gamma_{5}
     + C_{a3} k\cdot \epsilon \gamma_{5 }  \label{eq:Cterm} \\
   D_{a} &=& D_{a1} \eslash\gamma_{5} + D_{a2}  P\cdot \epsilon \gamma_{5}
     + D_{a3} k\cdot \epsilon \gamma_{5 } \ . \label{eq:Dterm}
\end{eqnarray}
After some algebra,  noting that $g^{\mu}_{\mu} = d = 4-2\varepsilon$ in the $d$-dimensional calculation, the coefficients $C_{ai}$ and $D_{ai}$ are found as
\begin{eqnarray}
   C_{a1}^{i} &=&   (\Imass +\Bmass)(1+\textfrac{\varepsilon}{2}) \\
   C_{a2}^{i} &=&   2 (3x-2 )+x \varepsilon\\
  C_{a3}^{i} &=&   2 (1-3y) -y \varepsilon  
\end{eqnarray}
and
\begin{eqnarray}
  D_{a1}^{i} &=&  0\\
   D_{a2}^{i} &=& 
   2(1-x) \left[ (x-y) y k^{2}- x(x-y) \Fmass^{2}
   \right.  \nn \\ &&
   + y (1-x) \Imass^{2}  + (y-1) \Bmass \Imass
   \nn \\ && \left.
    - (x-y) \Fmass  (\Imass+\Bmass) \right]  \\
   D_{a3}^{i} &=&  
    (2y -1) \left[ (x-y) y k^{2}  - x (x-y) \Fmass^{2}  
      \right. \nn \\ &&
+ y (1-x) \Imass^{2}+ (y-1) \Bmass \Imass
   \nn \\ && \left.
    -(x-y)\Fmass  (\Imass+\Bmass) \right]
\end{eqnarray}
After the $q$ integration, the divergent term is found only in the $\Amp_{1}$ amplitude as
\begin{equation}
   \xi_{a}^{i} = -\frac{Q_{M} B_{i} }{(4\pi)^{2}}\frac{1}{\varepsilon}\frac{\Imass+\Bmass}{2}\eslash \gamma_{5}
\end{equation}
The divergent terms in the $\Amp_{2}$ and $\Amp_{3}$ vanish due to 
$\int_{0}^{	1} dx \int_{0}^{x} dy (3x-2) =0$ and
$\int_{0}^{	1} dx \int_{0}^{x} dy (1-3y) =0$.
To calculate the divergent term we have used the following formula:
\begin{eqnarray}
\lefteqn{  \intqd \frac{q^{2}}{(q^{2}-S)^{3}} = \frac{i}{(4\pi)^{2-\epsilon}} \frac{4-2\epsilon}{2} \frac{\Gamma(\epsilon)}{\Gamma(3)}
  \left(\frac{1}{S}\right)^{\epsilon}} && \nonumber \\
  &=& \frac{i}{(4\pi)^{2}} \left( \frac{1}{\epsilon} - \log S-\frac{1}{2} - \gamma + \log 4 \pi +{\cal O}(\epsilon) \right) \label{eq:LoopS3q2}
\end{eqnarray}
%\com{The equation number in the first line has been taken away.}
The finite parts in the $\Amp_{2}$ and $\Amp_{3}$ {terms} are obtained as
\begin{widetext}
\begin{eqnarray}
  \Amp_{2a}^{i} &=&  \frac{Q_{M} B_{i}}{(4\pi)^{2}} \int_{0}^{1} dx \int_{0}^{x} dy 
  \left[
   2 (3x-2) \log S_{a}^{i} -x + \frac{D_{a2}}{S_{a}^{i}}
  \right] \\
  \Amp_{3a}^{i} &=&  \frac{Q_{M} B_{i}}{(4\pi)^{2}} \int_{0}^{1} dx \int_{0}^{x} dy 
  \left[
   2 (1- 3y) \log S_{a}^{i} + y + \frac{D_{a3}}{S_{a}^{i}}
  \right]
\end{eqnarray}
where the first two terms in the integrands are from the finite parts of the divergent integrals and the last terms come from the finite integrals.

Next, let us move to the calculation of the diagram (b). In a similar way, we obtain the finite $\Amp_{2}$ and $\Amp_{3}$ amplitudes as 
\begin{eqnarray}
  \Amp_{2b}^{i} &=&  \frac{Q_{B} B_{i}}{(4\pi)^{2}} \int_{0}^{1} dx \int_{0}^{x} dy 
  \left[
   2 (3y-1) \log S_{b}^{i} -y+1 + \frac{D_{b3}}{S_{b}^{i}}
  \right] \\
  \Amp_{3b}^{i} &=&  \frac{Q_{B} B_{i}}{(4\pi)^{2}} \int_{0}^{1} dx \int_{0}^{x} dy 
  \left[
   -2 ( 3y-1) \log S_{b}^{i} + y-1 + \frac{D_{b3}}{S_{b}^{i}}
  \right]
\end{eqnarray}
with $S_{b}$ defined in Eq.~(\ref{eq:defSb}) and 
\begin{eqnarray}
   D_{b2}^{i} &=&   2 ( y(1-y)(x-y) k^{2} - y(1-x)(x-y) \Fmass^{2}  -y^{2}(1-x) \Imass^{2} 
    \nn \\ && 
   +xy \Imass \Bmass + (x-y)((1-x) \Fmass+ \Bmass)( \Imass    +\Bmass))  \\
   D_{b3}^{i} &=&  2 (  -y(1-y)(x-y) k^{2} + y(1-x)(x-y) \Fmass^{2}  +y^{2}(1-x) \Imass^{2} 
   \nn \\ &&
    -xy \Imass \Bmass + y(x+y) ( \Imass+\Bmass)(\Fmass + \Imass) )
\end{eqnarray}
\end{widetext}
We have the divergent term in the $\Amp_{1}$ term as
\begin{equation}
   \xi_{b}^{i} = -\frac{Q_{B} B_{i} }{(4\pi)^{2}}\frac{1}{\varepsilon}\frac{\Imass-\Fmass-\Bmass}{2}\eslash \gamma_{5}\ .
\end{equation}

The amplitudes for diagrams (c), (d) and (e) have only the $\Amp_{1}$ 
components, which we do not use for the transition amplitudes. Only the 
divergent terms are necessary for the 
present arguments. 
The divergent terms are found
\begin{eqnarray}
  \xi_{c}^{i} &=&
   \frac{  Q_{M} B_{i}}{(4\pi)^{2}} \frac{1}{\varepsilon}
   \frac{\Fmass+2\Bmass}{2} \eslash\gamma_{5} \\
   \xi_{d}^{i} &=& -\frac{Q_{E} B_{i}}{(4\pi)^{2}} \frac{1}{\varepsilon} 
  \frac{(\Fmass^{2} - 2 \Bmass^{2} - 2 \Mmass^{2} + \Fmass \Bmass )}{2(\Imass+\Fmass)} \eslash \gamma_{5} \ \ \ \ \ \\
   \xi_{e}^{i} &=&  \frac{Q_{E}B_{i}}{(4\pi)^{2}} \frac{1}{\varepsilon} 
  \frac{ (\Imass^{2} - 2 \Bmass^{2} - 2 \Mmass^{2} -\Imass \Bmass )}{2(\Fmass+\Imass)} 
  \eslash \gamma_{5}\ .
\end{eqnarray}

At the end, collecting all the divergent terms of diagrams (a) to (e), we find that the divergent terms cancel according to
\begin{eqnarray}
 \lefteqn{ \sum_{A=a}^{e}\xi_{A}^{i} = } && \nn \\ &&
  -\left(\frac{B_{i} \eslash\gamma_{5}}{(4\pi)^{2}}\right)
   (Q_{E}-Q_{B}-Q_{M})    
  \frac{\Fmass - \Imass + \Bmass}{2}\frac{1}{\varepsilon}  \nn \\
  && =0
\end{eqnarray}
due to  the charge conservation $Q_{E}=Q_{B}+Q_{M}$. The cancellation takes place in each channel of the loop. 

\section{Results}
\begin{figure}[t]
\epsfxsize= 8.5cm
\begin{center}
  \epsfbox{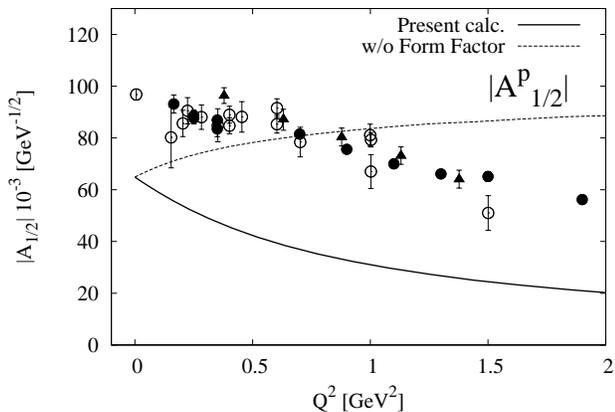}
\end{center}
\caption{Modulus of the $A_{1/2}$ helicity amplitude for the proton resonance as a function of $Q^{2}$ with $W=1535 \MeV$ calculated in the nonrelativistic formulation. The solid (dotted) line shows the $A_{1/2}^{p}$ amplitude with (without) the form factor of the meson inside the loops given in Eq.~(\ref{eq:FormFactor}).  
Marks with error bars are experimental data normalized by the $N^{*}$ full width $\Gamma_{N^{*}}=150$ MeV and the $N^{*} \rightarrow \eta N$ branching ratio $b_{\eta}=0.55$. Filled triangles and circles are results of the CLAS collaboration taken from Refs.\cite{Thompson:2000by} and  \cite{Denizli:2007tq}, respectively.  Open circles show results of Refs.\cite{Krusche:1995nv,Brasse:1977as,Beck:1974wd,Breuker:1978qr}. The values are taken from Ref.\cite{Krusche:1995nv}.
\label{fig:A1/2proton}}
\end{figure}

In this section we show our results for the helicity amplitudes, $A_{1/2}$ and $S_{1/2}$, of the $N(1535)$ dynamically generated in meson-baryon scattering. 
In Fig.\ref{fig:A1/2proton}, we show our result for the $A_{1/2}$ amplitude of the proton resonance calculated in the nonrelativistic formulation {[cf. Eqs. (\ref{eq:AampNR}, \ref{eq:amp2NR}, \ref{eq:amp3NR})]} with the CM energy $W=1535$ MeV. In the present calculation, 
we  multiply the amplitudes obtained in the former section by the electromagnetic form factors of the mesons or baryons to which the photon couples. The form factors of the meson and baryons components of the resonance, together with the intrinsic $Q^2$ structure of the loops are responsible here for the $Q^2$ dependence of the helicity transition form factors. For the mesons and baryons form factors we take monolope form factors consistent with the values for the radii of the mesons. We take
\begin{equation}
F(Q^2) = \frac{\Lambda^2}{\Lambda^2+ Q^2} \label{eq:FormFactor}
\end{equation}
with
\begin{eqnarray}
\Lambda_\pi &=& 0.727\ {\rm [GeV]} \\
\Lambda_K &=& 0.828\ {\rm [GeV]} 
\end{eqnarray}
which correspond to $\langle r^2 \rangle=0.44$ fm$^2$ and $\langle r^2 \rangle=0.34$ fm$^2$
for the pion and the kaon respectively \cite{Amendolia:1986wj,Amendolia:1986ui,Oller:2000ug}.
For the baryon form factor, we take the same form as for the corresponding meson to keep gauge invariance. 

In Fig.\ref{fig:A1/2proton}, we show our result for the $A_{1/2}^{p}$ amplitude 
of the proton resonance calculated in the nonrelativistic formulation together 
with various experimental data. The CM energy is 
taken as $W=1535$ MeV. 
Let us first discuss the $Q^{2}$ dependence of the helicity amplitude. 
The solid line denotes the modulus of the calculated amplitude multiplied 
by the meson form factor given in Eq.~(\ref{eq:FormFactor}), while the dotted 
line shows the results without the meson form factor, which means that
 the $Q^{2}$ dependence comes only from the loop calculation performed 
 in the previous section. {In this case,} the helicity amplitude increases as $Q^{2}$ increases. 
The inclusion of the form factors introduces a decreasing function of $Q^{2}$ which leads to a
$Q^{2}$ dependence of the helicity amplitude in fair agreement with the experimental observation, although it falls faster than experiment since at $Q^2=0$ we need a renormalization factor of 1.45 to reach the data while at $Q^2=1$ GeV$^2$ we need a factor of 2.15.

The absolute magnitude of our helicity amplitude looks underestimated if one 
 compares our result  directly with the experimental data shown in the figure. But it should be noted that extraction of the helicity amplitude from the experimental observables of the $\gamma p \rightarrow \eta p$ reaction is performed 
by using the following formula \cite{Trippe:1976aq,Krusche:1995nv,Armstrong:1998wg,Denizli:2007tq}:
\begin{equation}
   A_{1/2}(Q^{2}) = \sqrt{\frac{W \Gamma_{N^{*}}}{2m_{p} b_{\eta}} \sigma(W,Q^{2}})
\label{norma}
\end{equation}
with a  $N^{*}$ full width $\Gamma_{N^{*}}$, a $N^{*} \rightarrow \eta N$ branching ratio $b_{\eta}$, a resonance part of the total cross section $\sigma(W,Q^{2})$, the CM energy $W$ and  the proton mass $m_{p}$. 
%{ \bf
%0823
To obtain this relation, one assumes that the cross section is dominated by the single $N(1535)$ resonance and that the $S_{1/2}$ amplitude is small. 
%
%}
For the experimental data shown in Fig.\ref{fig:A1/2proton}, the amplitudes are normalized by $\Gamma_{N^{*}} = 150$ MeV and $b_{\eta}=0.55$ \cite{Denizli:2007tq}, which are obtained in a global fit of the cross section with the Breit-Wigner amplitude. 
On the other hand, the $N^{*}$ width obtained in the present approach is $\Gamma_{N^{*}} \simeq 74$ MeV for the $p^{*}$ as obtained from the pole position shown in Eq.~(\ref{eq:PPpstar}), in which the half width is given by the imaginary part. The branching ratio $b_{\eta}$ in this approach has been reported as $b_{\eta} \simeq 70$\%~\cite{inoue}. 
\begin{figure}[b]
\epsfxsize= 8.5cm
\begin{center}
  \epsfbox{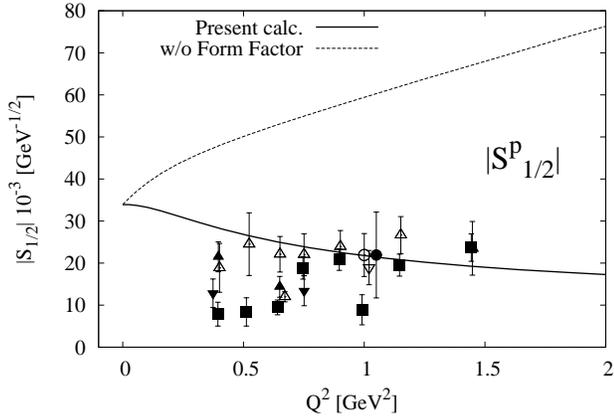}
\end{center}
\caption{Modulus of the $S_{1/2}$ helicity amplitude for the proton resonance as a function of $Q^{2}$ with $W=1535 \MeV$. The solid (dotted) line shows the $S_{1/2}^{p}$ amplitude with (without) the form factor of the meson inside the loops given in Eq.~(\ref{eq:FormFactor}). 
The sign of this amplitude  relative to $A_{1/2}^p$ is negative, both in experiment and theory. 
Solid triangles up (down) show the results from a combined analysis of $\pi$ ($\eta$) electroproduction data \cite{Aznauryan:2004jd,burkert}. The solid squares  are from single-$Q^2$ fits from Ref.~\cite{Tiator:2003uu}. The empty triangles up are taken from Ref.~\cite{Tiator:2006hd}. The other data (empty circle and empty triangle down) are from Ref. \cite{burkert}.
\label{fig:S1/2proton}}
\end{figure}
This normalization difference would give us a factor 1.6
reduction in the normalization with respect to the data shown in Fig.\ref{fig:A1/2proton}. Similarly, should one use in the experimental analysis a $N(1535)$ width of the order of 90 MeV as found at BES~\cite{Bai:2001ua} or the 100 MeV quoted in the last MAID2007 analysis \cite{Drechsel:2007if}, the results obtained would be in much better agreement with the theoretical results. 
We will come back to the discussion on the normalization later on when discussing the photoproduction cross section in the present approach. 
Let us note that the value obtained here {at $Q^2=0$ of $A_{1/2}^p=64.88\cdot 10^{-3}$ GeV$^{-1/2}$} is in excellent agreement with the most recent MAID 2007 analysis reported in Ref.\cite{Drechsel:2007if} of $66\cdot 10^{-3}\,\mbox{GeV}^{-1/2}$.

The $S_{1/2}^{p}$ amplitude calculated in the nonrelativistic formulation is plotted in Fig.\ref{fig:S1/2proton} together with experimental data. 
%\cor{
Although the modulus of $S_{1/2}$ is plotted in the figure, the ratio of $S_{1/2}$ to $A_{1/2}$ is nearly real and negative in agreement with experiment, where an implicit phase convention is taken that renders $A_{1/2}^{p}$ real and positive. 
Here we also show the effect of the meson form factor.

\begin{figure}
\epsfxsize= 8cm
  \epsfbox{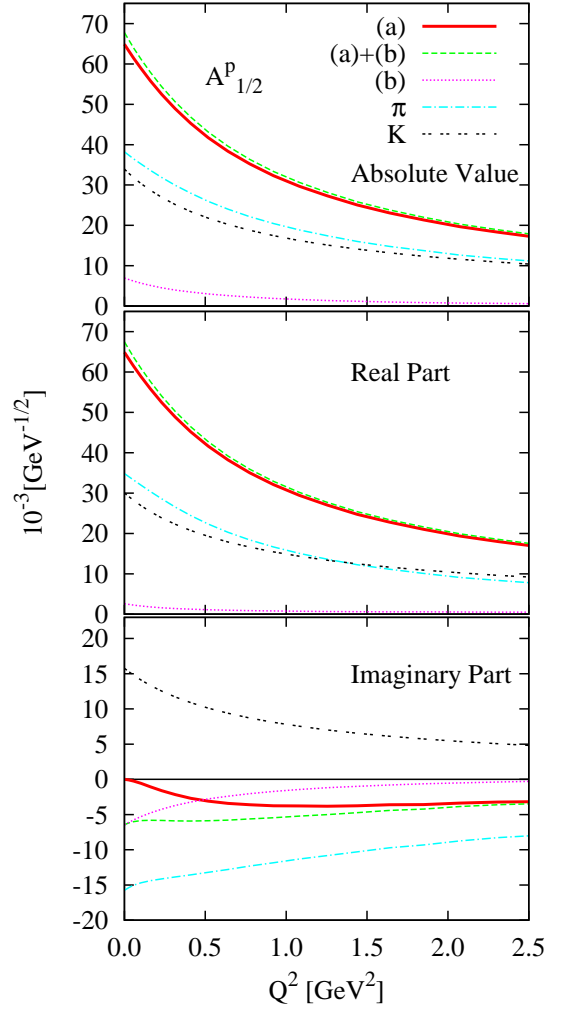} 
\caption{(Color Online) $A_{1/2}^{p}$ helicity amplitudes for the proton calculated 
in the nonrelativistic formulation as a function of $Q^{2}$. The upper, middle 
and lower panels are respectively the modulus, real parts and imaginary parts 
of the amplitudes. 
The phases of the amplitudes are set so that the $A_{1/2}^{p}$ amplitude has 
a real and positive value at $Q^{2}=0$.  
The solid lines show the calculation with diagram (a) (meson pole term). 
The dashed and dotted lines stand for the calculations of sum of diagrams (a) and (b) and diagram (b) only, respectively. The dot-dashed and two-dotted lines denote the pion and kaon contribution, respectively. \label{fig:protonA}  }
\end{figure}

\begin{figure}
\epsfxsize= 8cm
  \epsfbox{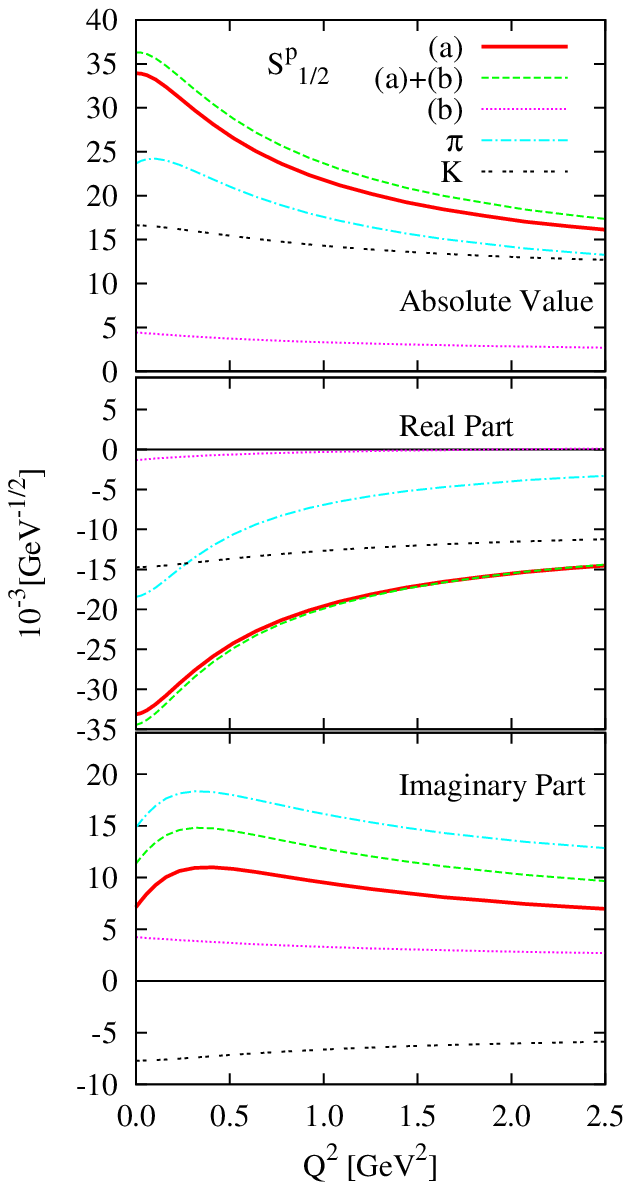}
\caption{(Color Online) $S_{1/2}^{p}$ helicity amplitudes for the proton calculated in the nonrelativistic formulation as a function of $Q^{2}$. Same as Fig.\ref{fig:protonA}.  \label{fig:protonS}}
\end{figure}

\begin{figure}
\epsfxsize= 8cm
  \epsfbox{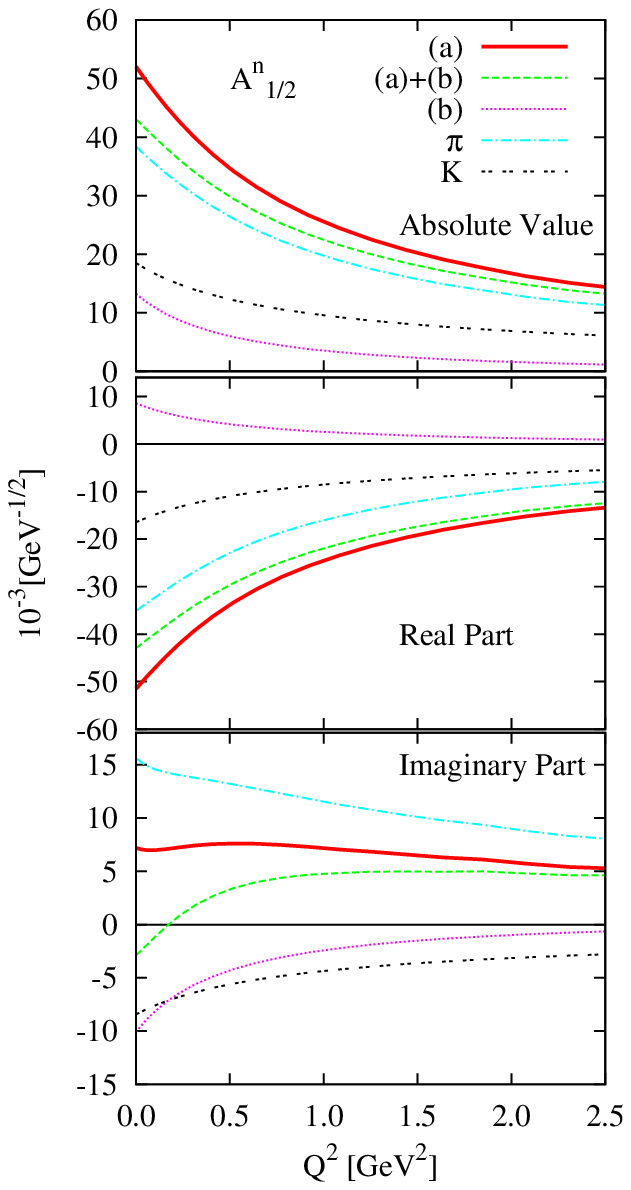} 
\caption{(Color Online) $A_{1/2}^{n}$ helicity amplitudes for the neutron calculated in the nonrelativistic formulation as a function of $Q^{2}$. The upper, middle and lower panels are respectively the modules, real parts and imaginary parts of the amplitudes. 
The phases of the amplitudes are set so that the $A_{1/2}^{p}$ amplitude has a real and positive value at $Q^{2}=0$.  
The solid shows the calculation with the diagram (a) (meson pole term). The dashed and dotted lines stands for the calculations of sum of diagrams (a) and (b) and diagram (b) only, respectively. The dot-dashed and two-dotted lines denote the pion and kaon contribution, respectively.  \label{fig:neutronA} }
\end{figure}

\begin{figure}
\epsfxsize= 8cm
  \epsfbox{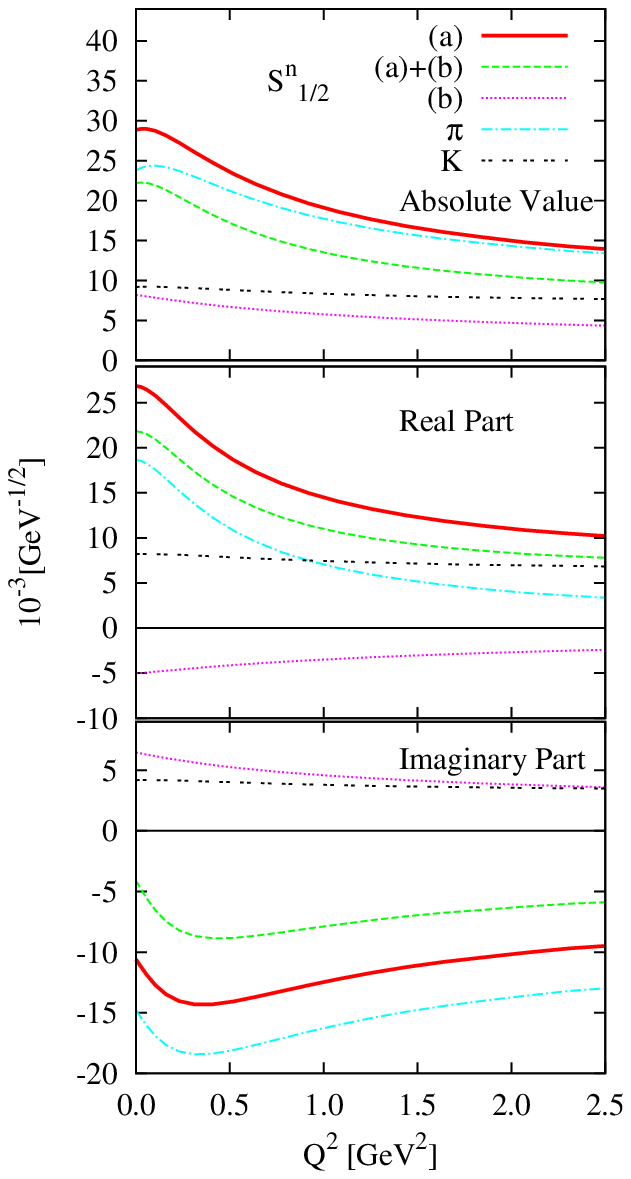}
\caption{(Color Online) $S_{1/2}^{n}$ helicity amplitudes for the neutron calculated in the nonrelativistic formulation as a function of $Q^{2}$. Same as Fig.\ref{fig:neutronA}.  \label{fig:neutronS} }
\end{figure}

As mentioned before, diagram (b), in which the photon couples to the baryon 
in the loop, gives sub-leading contributions in the $1/M$ expansion. This can
 be seen in Figs.\ref{fig:protonA}, \ref{fig:protonS}, \ref{fig:neutronA} and 
 \ref{fig:neutronS}. In these figures, the amplitudes with diagram (a) only 
 (solid line) are almost equivalent to those with both diagrams (a) and (b) 
 (dashed line), and {the contributions from} diagram (b)  (dotted line) are 
smaller, in the present case, than typical corrections of 20-30 percent for the $1/M$ terms.  
%very small. 
Therefore, the helicity amplitudes in the nonrelativistic formulation around these energies are basically given by diagram (a). We also plot the pion and kaon contributions separately. The figure shows that the pion contribution (dot-dashed line) is comparable with the kaon contribution (two-dotted line). This implies that the strange component is important in the structure of the $N(1535)$.

\begin{figure}
   \epsfbox{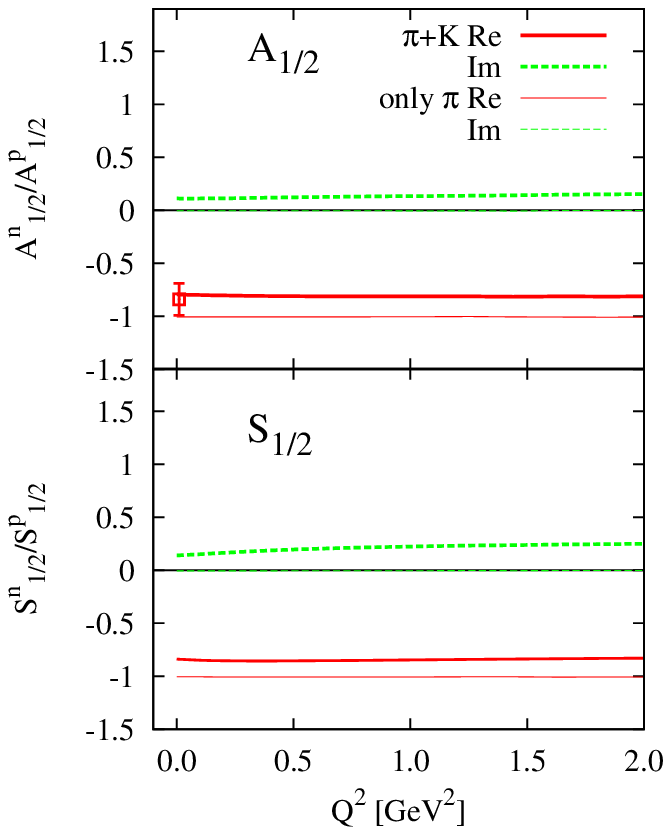}
   \caption{The $n/p$ ratios of the helicity amplitudes with $W=1535$ MeV. 
   The upper and lower panels show the $np$ ratios of the $A_{1/2}$ and $S_{1/2}$ amplitudes, respectively.
   The open square shows the value given in Ref.\cite{Mukhopadhyay:1995cr}.
      \label{fig:ratio}} 
\end{figure}

Next, we report on results for the helicity amplitudes of the neutron.
The $n/p$ ratios of the helicity amplitudes,  $A_{1/2}^{n}/A_{1/2}^{p}$ and $S_{1/2}^{n}/S_{1/2}^{p}$, are plotted in Fig.\ref{fig:ratio}
as a function of $Q^{2}$. 
For a real photon at $Q^{2}=0$ we obtain the ratio
$-0.79+ 0.11 i$, which is almost a real value, and its modulus $0.80$. 
A multipole analysis \cite{Mukhopadhyay:1995cr} using the inclusive experimental data of Ref.\cite{Krusche:1995nv} gives the negative sign value
$-0.84\pm0.15$ for $A_{1/2}^{n}/A_{1/2}^{p}$. 
Values of $|A_{1/2}^{n}|/|A_{1/2}^{p}|$ which are extracted from the ratio of the eta photoproduction cross sections, $\sigma_{n}/\sigma_{p}$, are reported as $0.82\pm0.04$ in Ref.\cite{HoffmannRothe:1997sv} and  $0.819\pm0.018$ in Ref.\cite{Weiss:2002tn}. 
The result obtained in our approach agrees with the experimental data in both sign and magnitude.
This comparison is free from the normalization uncertainty of Eq.~(\ref{norma}). 

The values of the $A_{1/2}$ helicity amplitude at $Q^{2}=0$ are summarized in Table \ref{tab:AmpQ0}. The phases of the amplitudes are set so that the $A_{1/2}$ helicity amplitude for $p^{*}$ has a real and positive value at $Q^{2}=0$. The ratios of the helicity amplitudes to that of $p^{*}$ are also shown in the table. We also show the helicity amplitudes in the isospin decomposition:
\begin{eqnarray}
   A_{1/2}^{\rm IS}  &=& \frac{1}{2}  \left( A_{1/2}^{p} + A_{1/2}^{n} \right) \\
   A_{1/2}^{\rm IV}  &=& \frac{1}{2}  \left( A_{1/2}^{p} - A_{1/2}^{n} \right).
\end{eqnarray}
In the nonrelativistic calculation we find that the value of the isoscalar component is much smaller than that of the isovector, which is consistent with experimental observation.  The ratio of the isoscalar component to the $p^{*}$ amplitude is $|A_{1/2}^{\rm IS}/A_{1/2}^{p}|=0.12$ in our calculation, while in experiments it is found to be $A_{1/2}^{\rm IS}/A_{1/2}^{p} = 0.09 \pm 0.02$ in Ref.\cite{HoffmannRothe:1997sv} and $0.09\pm 0.01$ in Ref.\cite{Weiss:2002tn}. 

{It is also interesting to compare the values of our $p^*$ and $n^*$ helicity amplitudes $A_{1/2}$ at $Q^2=0$ with those of the PDG \cite{Yao:2006px}. We obtain $0.065$ GeV$^{-1/2}$ and $-0.052$ GeV$^{-1/2}$ for the $p^*$ and $n^*$, respectively, versus the values quoted in the PDG, which include uncertainties from the compilation of data of several analyses, $0.090\pm 0.030$ GeV$^{-1/2}$ for the $p^*$ and $-0.046\pm 0.027$ GeV$^{-1/2}$ for the $n^*$. As one can see, the agreement, within uncertainties, is good.}
\begin{table}
\begin{tabular}{ccccc}
\hline \hline
\multicolumn{5}{c}{Nonrelativistic calculation}\\
\hline
  & $A_{1/2}$ & $|A_{1/2}|$ & $A_{1/2}/A_{1/2}^{p}$ & $|A_{1/2}/A_{1/2}^{p}|$\\
\hline
$p^{*}$  &  $64.88 $ &   $64.88 $ & --- & --- \\
$n^{*}$ & $ -51.54 + 7.21 i $ & $ 52.04 $  & $ -0.79 +  0.11 i$  & $ 0.80$ \\
IV & $ 58.21-3.61 i$ &  $58.32 $ &   $ 0.90 -0.056 i$   & $0.90$ \\
IS & $ 6.67 +  3.61 i$ &  $ 7.59 $ &  $ 0.10 +  0.056 i$ &   $0.12$ \\
\hline\hline
\multicolumn{5}{c}{Relativistic calculation}\\
\hline
  & $A_{1/2}$ & $|A_{1/2}|$ & $A_{1/2}/A_{1/2}^{p}$ & $|A_{1/2}/A_{1/2}^{p}|$\\
\hline
$p^{*}$  &  $46.31 $ &   $46.31 $ & --- & --- \\
$n^{*}$ & $ -55.24 + 28.95 i $ & $ 62.36 $  & $ -1.19 +  0.63  i$  & $ 1.35$ \\
IV & $ 50.78-14.47 i$ &  $52.80 $ &   $ 1.10 -0.31 i$   & $1.14$ \\
IS & $ -4.46 +  14.47 i$ &  $ 15.15 $ &  $ -0.10-  0.31  i$ &   $0.33$ \\
\hline\hline
\end{tabular}
\caption{
Values of the $A_{1/2}$ helicity amplitudes at $Q^{2}=0$ in units of $10^{-3}$GeV$^{-1/2}$ in the nonrelativistic (upper panel) and relativistic (lower panel) calculations. The ratios to the $A_{1/2}^{p}$ are also shown. The phases of the amplitudes are set so that the $A_{1/2}^{p}$ amplitude has a real and positive value at $Q^{2}=0$. IV and IS stand for isovector and isoscalar. 
\label{tab:AmpQ0}}
\end{table}

\section{Discussion}

\subsection{Photoproduction of the eta meson}
\label{sec:defallphoto}
In this section, we investigate the photoproduction of the eta meson close to threshold energies in order to discuss the normalization of the helicity amplitude in the present approach. 
In the calculation of the helicity amplitude in Sec. \ref{sec:norela}, we have separated out the $N(1535)$ resonance contributions from the scattering amplitudes in which the $N(1535)$ is dynamically generated, by setting the CM energy as the resonance energy and multiplying  the coupling strengths of the $N(1535)$ resonance to each channel, $g_{N^{*}}^{i}$,  by the loop functions.  
For our purpose of calculating the $\eta$ photoproduction cross section,  we replace the coupling strengths, $g_{N^{*}}^{i}$, by the $MB \rightarrow \eta p$ scattering amplitudes, $t^{(i)}_{\eta p}$, obtained by the chiral unitary approach \cite{inoue}, where $i$ denotes the initial meson baryon channel.
The evaluation of these amplitudes is sketched here in section \ref{sec:formulation} and done in detail in Ref.\cite{inoue}. 

Following the above prescription for the eta photoproduction amplitudes, 
we obtain the cross section of the photoproduction as 
\be
\sigma=\frac{M^2}{4\pi\;s}\;\frac{k_\eta}{k_\gamma}|t_{\gamma p\to\eta p}|^2
\ee
where $k_\gamma$ ($k_\eta$) is the photon ($\eta$)  three-momentum in the CM frame and the $T$-matrix is given by 
\be
|t_{\gamma p\to\eta p}|^2=\frac{8m_{p}^2+8EE^{\prime}}{16m_{p}^2}\; \left|\sum_{i=1}^6{\cal M}_1^{i \rm (NR)}\frac{t^{(i)}_{\eta p}}{g_{N^{*}}^{i}}\right|^{2} 
\label{tdef}
\ee
where $E$ ($E^{\prime}$) are the energies of the incoming (outgoing) proton in the CM frame. To obtain the $\Amp^{i \rm (NR)}_{1}$ amplitude from the $\Amp^{i \rm (NR)}_{2}$ and $\Amp^{i \rm (NR)}_{3}$ amplitudes calculated in the previous section, we use the gauge invariance condition given in Eq.~(\ref{eq:gauseinvNR}). Actually, the eta photoproduction with a real photon has no contribution from the $\Amp^{i \rm (NR)}_{3}$ amplitude, hence only one amplitude needs to be evaluated which we choose to be $\Amp_{2}$ that shows its finiteness immediately.

\begin{figure}
\begin{center}
\epsfxsize= 8.5cm
\epsfbox{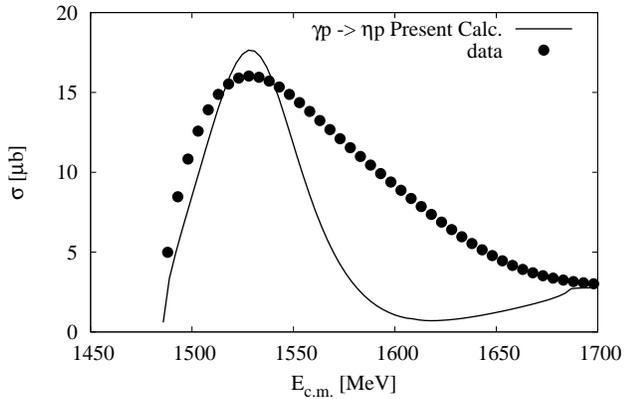}
\end{center}
\caption{Cross section for photoproduction of the $\eta$ using the nonrelativistic formalism for the photon loop. Dots: Analysis from Arndt \cite{Arndt:2003if} (SAID data base). 
}
\label{fig:total_photo_NR}
\end{figure}

In Fig.\ref{fig:total_photo_NR}, 
the total cross section of the present calculation (solid line) is plotted together with the data from Arndt \cite{Arndt:2003if} (dots). Our result of the eta photoproduction cross section 
provides the right strength around the peak of the $N^*(1535)$ resonance but the
width of the peak is narrower than the experiment as a result of the narrower widths of the $N(1535)$ resonance obtained by the present model. 

\begin{figure}[h]
\begin{center}
\epsfxsize= 8.5cm
\epsfbox{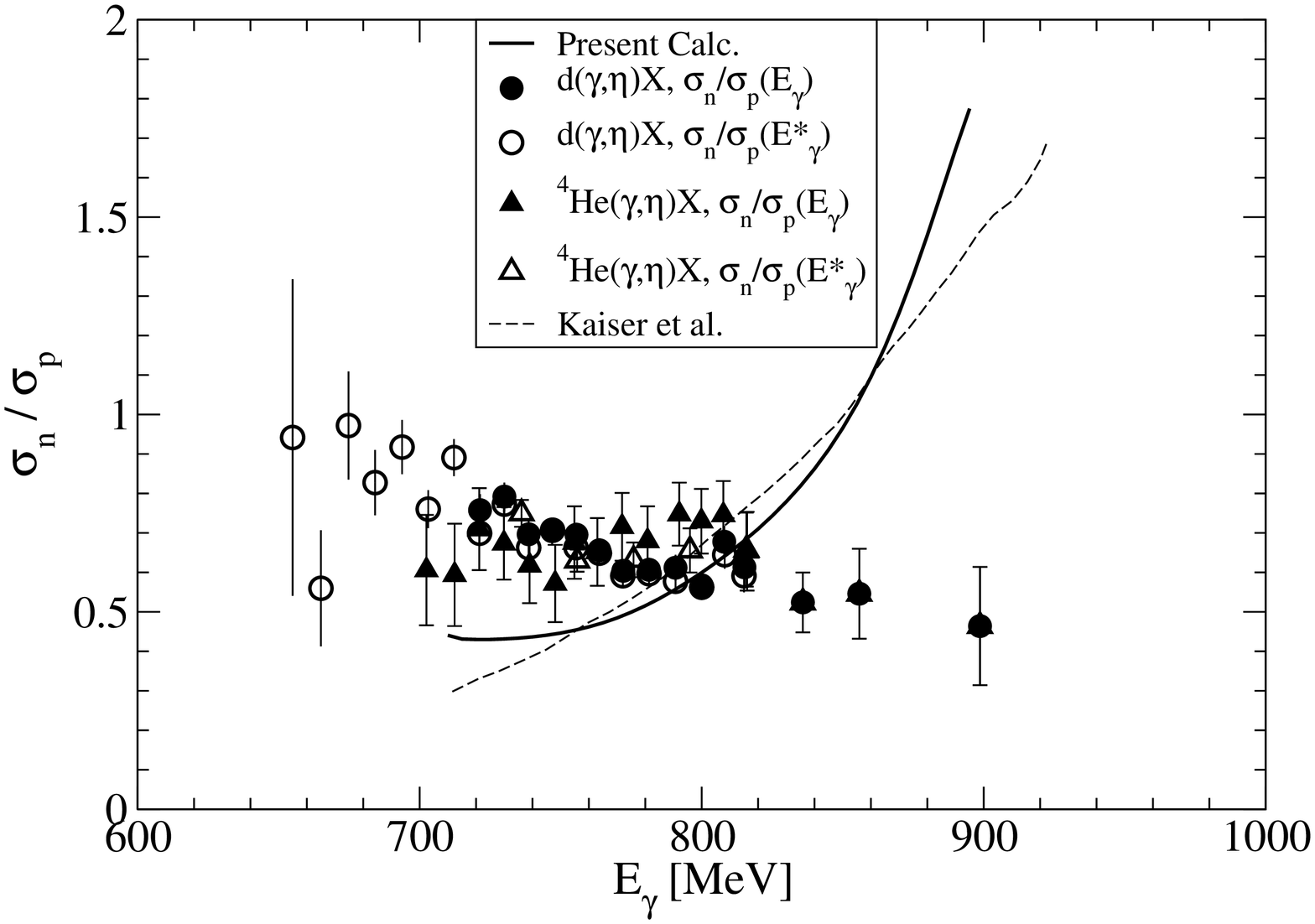}
\end{center}
\caption{
Ratio of cross sections of photoproduction on the neutron over that on the proton, $\sigma_n/\sigma_p$,  as a function of the photon energy $E_\gamma$ in the laboratory frame. The data are taken from Ref.\cite{Weiss:2002tn} for the deuteron target and Ref.\cite{Hejny:1999iw} for the Helium target. The dashed line is a theoretical calculation by Kaiser {\it et al.} in Ref.\cite{kaiser}. 
}
\label{fig:ratioNR}
\end{figure}

In Fig.\ref{fig:ratioNR} we show the ratio of the cross sections of $\eta$ 
photoproduction on the neutron over that on the proton, $\sigma_n/\sigma_p$, obtained in the nonrelativistic formulation, comparing our calculation with experimental data. 
The value of the ratio at $E_{\gamma}^{\rm lab}=785$ MeV which corresponds to $E_{\rm CM} = 1535$ MeV is found to be 0.53, which is quite consistent with the experimental data.

Although we are only concerned with the vicinity of the $N^*(1535)$ resonance, 
one cannot overlook the apparent discrepancy of the theory and experiment at 
photon energies above 800 MeV as shown in Fig. \ref{fig:ratioNR}, which is also 
shared by the model of Ref. \cite{kaiser}. Only very recently have we obtained 
experimental information that brings a new perspective to these discrepancies. Indeed, in Ref. \cite{jaegle_thesis} the theoretical results of Fig. \ref{fig:ratioNR} are taken and folded with the Fermi motion of the nucleons in the deuteron to allow a realistic comparison with the experimental data. In \cite{jaegle_thesis} it is shown that the steep rise of the theoretical curve is softened to a curve in between the one of Fig. \ref{fig:ratioNR} and a horizontal line starting from 800 MeV. On the other hand, recent results from \cite{Krusche_preli} for $\gamma n\to\eta n$ show a steady rise starting from $E_\gamma=900$ MeV. These two facts together would render the apparent discrepancies into a rough qualitative agreement. Let us mention in this respect that the inclusion of the $\pi\pi N$ channel, although only qualitatively considered as shown in the next section, also works in the direction of softening the steep rise of Fig. \ref{fig:ratioNR}. In any case, we must admit larger theoretical uncertainties at higher energies than around the resonance region, also including extra terms considered in Ref. \cite{Borasoy:2007ku} that would become relevant as one moves away from the resonance pole.

\subsection{Higher order couplings}
\label{sec:hiorder}
In the nonrelativistic treatment of the photon loops from Fig. \ref{fig:loops} in Sec. \ref{sec:norela}, the magnetic couplings of the photon to the baryons have been neglected as they are of higher order in the external photon momentum according to $k/M$. The convection part of the $\gamma BB$ coupling shows a similar $p/M$ suppression, where $p$ is a typical loop momentum. These higher order terms have been neglected for the sake of consistency with the hadronic part of the model: as discussed after Eq.~(\ref{propnorela}), only the positive energy part of the baryon propagator is taken in the evaluation of the $MB\to MB$ scattering amplitude. In this section, the effects of the higher order $\gamma BB$ coupling are studied, which can give an idea of theoretical uncertainties from these terms.

The photon-baryon coupling is given by 
\ba
{\cal L}_{\gamma BB}&=&-\overline{\Psi}\left(Q_{B}\Aslash\,+\frac{\kappa e}{2M_{N}}\,\sigma^{\mu\nu}\partial_\mu A_\nu\right)\Psi
\ea
with $\sigma^{\mu\nu}=\frac{i}{2}[\gamma^\mu,\gamma^\nu]$, 
the baryon charge $Q_{B}$ and the anomalous magnetic moment $\kappa$ given in units of the nuclear magneton $\mu_N=e/(2M_N)$.
In the nonrelativistic reduction of this interaction, only terms up to order $p/(2M)$ are considered which leads to the vertex
\ba
-it&=&\frac{iQ_{B}\,F(Q^2)}{2M}\,\vec{\epsilon}\,(\vec{p}+\vec{p'})+\frac{e\,G_M(Q^2)}{2M_{N}}\,\vec{\epsilon}\,(\vec{k}\times\vec{\sigma}) \mu_{B} \non
\label{vertexmag}
\ea
where we have supplied the form factor $F(Q^{2})$ given in Eq.~(\ref{eq:FormFactor}) and the Sachs form factor $\displaystyle G_M(Q^2)=1/(1+Q^2/\Lambda_M^2)^2$, $\Lambda_M^2=0.71$ GeV$^{2}$
and $\mu_{B}$ is the baryon magnetic moments in units of $\mu_N$ from the PDG \cite{Yao:2006px}. We use common form factors $G_{M}$ for the $\Sigma$ and $\Lambda$.

With the vertex from Eq.~(\ref{vertexmag}), diagram (b) from Fig.\ref{fig:loops} can be calculated. For the convection part, the result has been already obtained in Eq.~(\ref{convection}). 
The magnetic part of diagram (b) is given by
\begin{eqnarray}
t_{{\rm mag}}^{(i)}&=&-i\,e\;\mu_B^{(i)}\frac{g^{i}_{A}\,g_{N^{*}}^{i}}{2f}\,\vec{\eee}\,(\vec{\sigma}\times {\vec k})\int\frac{d^4 q}{(2\pi)^4}\;\frac{\vec{\sigma}(\vec{P}-\vec{q})}{q^2-M_{i}^2}\non
&\times&\frac{2M_{i}}{(q-k)^2-M_{i}^2}\;\frac{1}{(P-q)^2-m_{i}^2}.
\label{pimp}
\end{eqnarray}
This expression is finite and not logarithmically divergent as the convection part of the $\gamma BB$ coupling or diagrams (a) and (c). The magnetic part is gauge invariant by itself as the structure of Eq.~(\ref{pimp}) shows. In Eq.~(\ref{pimp}), $m(M)$ are the masses of the meson (baryon) of channel $(i)$, $P^2\equiv s$, and $g_{N^{*}}^{i}$ are the coupling strengths to the $N(1535)$ from Tables \ref{tab:ncoupl} and \ref{tab:pcoupl}. For photoproduction, one obtains the amplitude $T(\gamma N\to\eta N)$ by replacing $g_{N^{*}}^{i}$ with the $MB \rightarrow N\eta$ $T$-matrix as discussed in Sec.\ref{sec:defallphoto}, summing over all channels $i$. The axial charges $g_{A}^{i}$ are given in Table \ref{tab:gA}. We also take into account the magnetic $\Sigma^{0}\Lambda$ transition. We choose a negative $\mu_{\Sigma^0\Lambda}=-1.61$ which is the prediction of the quark model \cite{Aliev:2001uq} while only the modulus can be measured \cite{Yao:2006px}. For the photon loops with a $\Sigma^0\Lambda$ transition, we use average masses for the baryons.
For the unknown magnetic moment of the $\Sigma^0$, we take $\mu_{\Sigma^{0}}=\frac{1}{2}(\mu_{\Sigma^{+}}+{\mu_{\Sigma^{-}}})=0.65 $ which is obtained by the SU(3) argument \cite{Coleman:1961jn} and also consistent with the quark model.

Evaluating the loop integral using Feynman parameters we obtain 
\begin{equation}
t_{\rm  mag}^{(i)}=-ih^{(i)}\left({\vec k}^{\,2} \vec{\sigma}\cdot \vec{\eee}-\vec k\cdot\vec{\sigma}\;\vec k\cdot \vec{\eee}\right)
\end{equation}
with 
\begin{equation}
h^{(i)}= -e \mu_{B}^{(i)} \frac{g_{N^{*}}^{i}\,g_{A}^{i}\,M_i}{16\pi^2\,f} \int\limits_0^1 dx\int\limits_0^xdy\;\frac{y}{S_{b}^{i}-i\epsilon}.
\end{equation}
where $S_b^i$ is defined in Eq.~(\ref{eq:defSb}) and we used $P\cdot \sigma =0 $ in the CM frame. 
After summing up all channels for the photon loop, we obtain the contribution to the helicity amplitudes from the magnetic couplings according to
\begin{eqnarray}
A^{1/2}_{\rm mag}&=&-\sqrt{\frac{2\pi\alpha}{q_R}}\,\frac{\sqrt{2}\,{\vec k}^{\, 2}}{e} \sum_{i=1} h^{(i)} \\
S^{1/2}_{\rm mag}&=&0.
\end{eqnarray}
The $\gamma BB$ coupling of diagram (b) from Fig.\ref{fig:loops} has a convection part and a magnetic part (cf. Eq.~(\ref{vertexmag})), and both are of order $1/M$. Thus, we will treat both parts together 
and compare them to the previous results when only the leading order 
couplings are considered. The latter appear in diagrams (a) and (c), while in the nonrelativistic calculation diagrams (d) and (e) do not contribute.

In order to see the effects of the various $1/M$ terms, we have compared them 
in Tab. \ref{tab:mag_cont}. The phase is chosen in the way that 
$A_{1/2}^p (Q^2=0)$, including all $1/M$ terms, is real and positive.
\begin{table}
\caption{The $1/M$ contributions from the baryon pole term for $A_{1/2} (Q^2=0)\, [10^{-3} \mbox{GeV}^{-1/2}]$. See text for the different cases.}
\begin{center}
\begin{tabular}{lll}
 \hline\hline
\hspace*{0.7cm}&$A_{1/2}^p$\hspace*{1.2cm}&$A_{1/2}^n$\\
I&$ 64.7+4 i$&$-51.9+4 i$\\
II&$67.7-2 i$&$-42.7-5.5 i$\\
III&$70.0 -2.7 i$&$-39.9 + 4.6 i$\\
IV&$74.7$&$ -44.5+1.9 i$\\
 \hline\hline
\end{tabular}
\end{center}
\label{tab:mag_cont}
\end{table}
Case (I) is from diagrams (a) and (c) of Fig. \ref{fig:loops} only, without 
any $\gamma BB$ couplings. Case (II) also includes the convection part of the 
$\gamma BB$ coupling from diagram (b). Case (III) additionally includes the magnetic part. Case (IV) includes, on top of the other contributions, the $\Sigma^0\Lambda$ transition magnetic part. 

From the different contributions, we can see that the $\gamma BB$ coupling has moderate influence on the results: For $A_{1/2}^p$, the result increases due to the convection part and the magnetic part. For $A_{1/2}^n$, the various $1/M$ contributions lead to a decrease as Table \ref{tab:mag_cont} shows. As a result, the ratio $A_{1/2}^n/A_{1/2}^p=-0.60 + i\,0.03$ is smaller than the value of $-0.79+i\,0.11$ found in the last section, where only the leading couplings were included. 

\begin{figure}
\epsfxsize= 8.5cm
\begin{center}
  \epsfbox{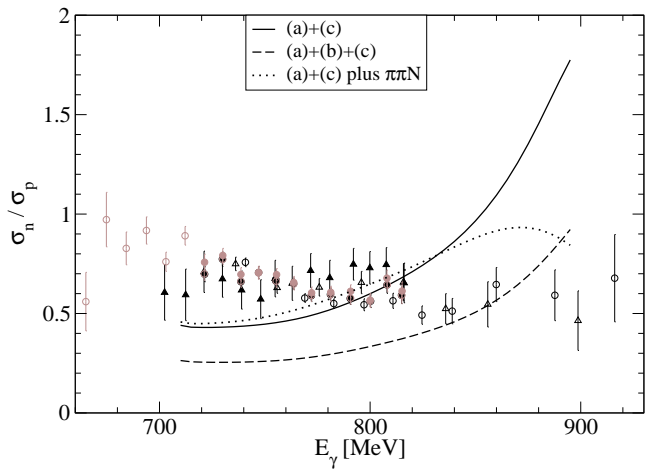}
\end{center}
  \caption{(Color online) The ratio $\sigma_n/\sigma_p$ including the baryon pole diagram (b) (dashed line), compared to the result without diagram (b) (solid line). Also, the result is shown when the $\pi\pi$N channel from \cite{inoue} is included. {The data is from Refs. \cite{Weiss:2002tn,Hejny:1999iw} [see also Ref. \cite{Krusche:2003ik}].} }
  \label{fig:snsp_mag}
\end{figure}

In Fig. \ref{fig:snsp_mag}
the ratio of $\eta$ photoproduction on the neutron over that on the proton is
shown. The decrease, when including the higher order couplings of diagram (b),
reflects the results from Table \ref{tab:mag_cont}, that $A_{1/2}^p$ increases
and $A_{1/2}^n$ decreases when including the $1/M$ corrections. In the same
figure, we show the result when including additional ingredients for the
rescattering model from \cite{inoue} (dotted line). These are the $\pi\pi N$
channel in rescattering as well as a form factor for the $MB\to MB$ transitions.
Coupling the photon to these ingredients is beyond the scope of this work;
therefore, they are not included in the final results. Including these
ingredients in the rescattering part, the ratio drops for higher photon
energies, and this gives an idea of theoretical uncertainties coming from omitting the $\pi\pi N$ channel in the present study. For other observables discussed in this study, the additional ingredients from \cite{inoue} lead only to very minor changes.

For consistency with the approach followed for meson baryon scattering, 
the terms of order $1/M$ should be omitted, and the results of the former 
section should be used to compare with data. The discussion in this section 
gives us an idea of the uncertainties that one may have when considering 
the $1/M$ terms.

\subsection{Result of the relativistic formulation} 
		
\begin{figure}[h]
\epsfxsize= 8.5cm
\begin{center}
  \epsfbox{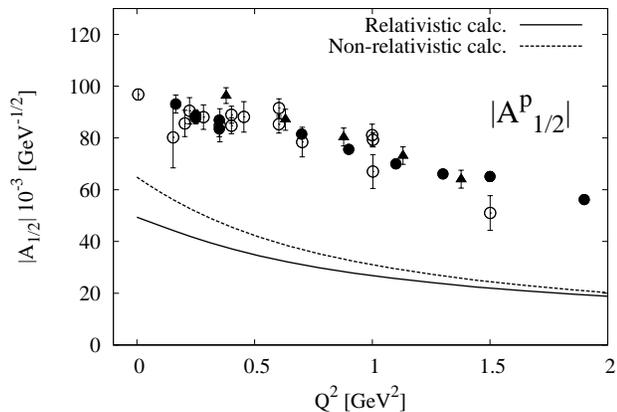}
\end{center}
\caption{Modulus of the $A_{1/2}$ helicity amplitude for the proton resonance as a function of $Q^{2}$ with $W=1535 \MeV$ in relativistic formulation. 
The solid (dotted) line shows the $A_{1/2}^{p}$ amplitude calculated in the relativistic
(nonrelativistic) formulation.  
The marks are same as in Fig.\ref{fig:A1/2proton}.
\label{fig:relaA1/2}}
\end{figure}

\begin{figure}[h]
\epsfxsize= 8.5cm
\begin{center}
  \epsfbox{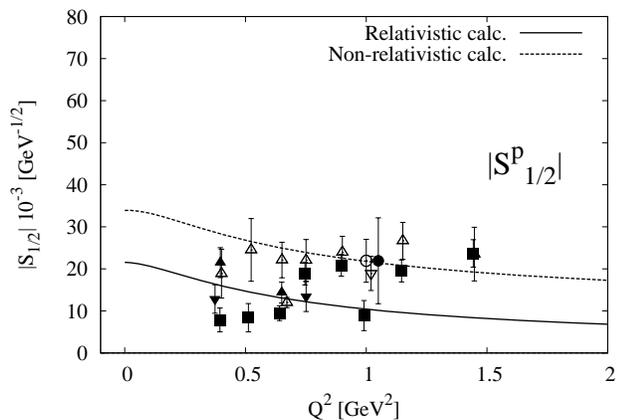}
\end{center}
\caption{Modulus of the $S_{1/2}$ helicity amplitude for the proton resonance as a function of $Q^{2}$ with $W=1535 \MeV$ in relativistic formulation. 
The solid (dotted) line shows the $S_{1/2}^{p}$ amplitude calculated in the relativistic
(nonrelativistic) formulation. 
The marks are same as in Fig.\ref{fig:S1/2proton}.
\label{fig:relaS1/2}}
\end{figure}

Here we briefly discuss our result for the $A_{1/2}^{p}$ amplitude of the $p^{*}$ resonance calculated in the relativistic formulation given in Sec.\ref{sec:relacal}. As we have already mentioned, the relativistic formulation is less consistent with the model of the $N(1535)$ resonance generated  dynamically in the present approach than the nonrelativistic formulation. Therefore we rely upon the nonrelativistic calculation. 
However, the $Q^{2}$ dependence could be given better by the relativistic calculation, particularly if we go to values of $Q^{2}$ of the order of 1 GeV$^{2}$ or above. 
On the other hand, in the relativistic formulation we have found the cancellation of divergences coming from each diagram. In Fig.\ref{fig:relaA1/2} we show the results for the $A_{1/2}^{p}$ amplitude in the relativistic calculation shown by the thick solid line in comparison with the nonrelativistic calculation shown by the dashed line. The relativistic result is a bit below the nonrelativistic calculation. 
We also plot the results for the $S_{1/2}^{p}$ amplitude obtained in the relativistic formulation in Fig.~\ref{fig:relaS1/2}. 
The relativistic calculation gives a smaller result than the nonrelativistic one, as in the case of the $A^{p}_{1/2}$ amplitude, but the differences are now larger. This reflects the fact that the $S^{p}_{1/2}$ amplitude is more sensitive to small changes of the input and, consequently, one must accept larger theoretical uncertainties in this amplitude. The dispersion of the data seems to reflect a similar problem on the experimental side, the results proving also rather sensitive to the assumptions made in the different analyses.
%
%Again the relativistic calculation underestimates the absolute values of the amplitude.
%
%We also show the relativistic amplitude multiplied by a factor 2.3 in order to compare the $Q^{2}$ dependence of our relativistic amplitude to the experimental data.

   Although we  prefer the nonrelativistic results of $A_{1/2}^{p}$  at $Q^2 =0$, as
   already mentioned, one can take this difference as a measure of the
   theoretical uncertainties.
   At $Q^2=0$ the relativistic result is some 25\% below the non relativistic one.
   Taking into account 10\% of uncertainty in the recent MAID2007 analyis 
   \cite{TiatorPC} of $A_{1/2}^p=66\pm 7\cdot 10^{-3}\mbox{GeV}^{-1/2}$, one finds good
    agreement of the theory with the MAID2007 results within uncertainties.

%0829

{Let us note that in the relativistic calculation the factor needed to agree with data at $Q^2=0$ is 1.9 and at $Q^2=1$ GeV$^{2}$ the factor needed is 2.4. It indicates a faster fall down than experiment, 25 \% lower than exeriment at $Q^2=1$ GeV$^{2}$ with a curve normalized at $Q^2=0$. This compares with 48 \% smaller strength than experiment at $Q^2=1$ GeV$^{2}$ of the nonrelativistic curve, normalized to the data at $Q^2=0$. This indicates that relativistic effects play some role at large $Q^2$, as one might think, along the line of similar findings in relativistic quark models \cite{capstick,metsch}.}

We have also calculated the transition amplitudes using $N^{*} BM$ couplings with a derivative
of the type $\gamma_\mu \partial^\mu$
in the relativistic calculation. In this case, we have an extra momentum in the $N^{*} BM$ coupling and this momentum is included in the loop integral. We also have an extra Kroll-Ruderman contact term. As a consequence,  the cancellation of the divergences is not complete in the case of $Q^{2}>0$, while, for the real photon, that is $Q^{2}=0$, 
the sum of the amplitudes is finite.
%the amplitudes are finite without  divergences. 
From the viewpoint of consistency with the model of the $N^{*}$, one should not
use the derivative coupling in the $N^{*}BM $ vertex. In the unitarization based
on the $N/D$ method, we exploit the so-called elastic unitarity, in which the
interactions $V$ are evaluated on the mass shell. In the present case, the
momenta in the Weinberg-Tomozawa couplings are set on the mass shell. Therefore,
to maintain consistency with this procedure, the momentum in the $N^{*}BM$ should have the value on the mass shell and should not be included in the loop integral. This means that a constant $N^{*}BM$ coupling is more consistent with the present $N^{*}$ model. 
In any case, just for illustrative purposes, the value that we obtain for $A^p_{1/2}$ with the off shell derivative coupling is of the order of $90\cdot 10^{-3}$ GeV$^{-1/2}$.
Our, preferred, nonrelativistic result lies in between these two illustrative
relativistic results.

\section{Summary and critical observations}

In this work we have addressed the evaluation of the electromagnetic helicity
form factors for the electroproduction of the $N^*(1535)$ resonance considered
as a dynamically generated resonance.  For this purpose the coupling of the
photon to the meson baryon components of the N(1535), previously studied within the
chiral unitary  approach to pion nucleon scattering, was considered. The calculations have been
done relativistically and nonrelativistically, and both of them are found to lead to
finite results for the transition amplitudes, as well as for eta photoproduction
which is evaluated simultaneously with the same formalism.

Our study finds interesting results 
which we summarize here:

The amplitudes were obtained without any free parameters, 
since the couplings of the resonance to the channels have been 
obtained from a previous study of $\pi N$ scattering.  
The agreement with the $A_{1/2}^{p}$ amplitude of
the proton $N^*(1535)$ resonance 
is fair up to the normalization problem that we have discussed. Indeed, we showed
that the absolute values of the experimental
amplitudes were tied to assumptions on the total width of the resonance, which
is still far from being a settled issue. 
We also showed that our results for $A_{1/2}$ {at $Q^2=0$} are in perfect agreement with the
most recent MAID2007 analysis of scattering and photoproduction data. 
The $Q^2$ dependence of the transition
form factor obtained was in fair agreement with the experimental determination, although it provided a moderately faster fall down than experiment.
This result is by no means obvious within the picture of a dynamically
generated resonance, since the $Q^2$ dependence should be provided
by the meson form factors and they fall much faster than these experimental form
factors. Yet, we found that the theory, in the absence of the meson form
factors,
provided a rising function of $Q^2$, due to the structure of the loops involved,
which led to a moderate decrease of the $N^*(1535)$ transition form factors
when the meson form factors were considered.

  The results obtained for the $S_{1/2}$ amplitude  are also in fair agreement
with experiment, both in size and the relative sign to the  $A_{1/2}$
amplitude. It should be stressed that the nature of the loops, where some
intermediate states can be put on shell, naturally leads to an imaginary part
of the amplitude and hence one obtains complex transition form factors.
Comparison with the data implies a choice of phase to make the  $A_{1/2}$
amplitude real and with the sign chosen in the experimental analysis. However,
once this is done, the rest of the amplitudes have very well
determined signs and phases . In this sense we found that the ratio of the $S_{1/2}$ to the
$A_{1/2}$ amplitude was practically real and negative, and we also found that
the ratio of the $A_{1/2}$ amplitude of the neutron resonance to that of the
proton resonance was also practically real and of the order of -0.80, in good
agreement with experiment.

It should be noted that the signs and strengths of the different amplitudes are a
nontrivial consequence of the contribution of the different channels in the
photon transition loops and of subtle interference of terms.

    Thus we can say, that 
the agreement with the data is fair when it comes to the shape of
the $Q^2$ dependence and good in the ratios
of amplitudes which are free of the global normalization. 
All these features together provide a boost to the hypothesis of
the $N^*(1535)$ as being a dynamically generated resonance. This of course does
not exclude some other components beyond those of meson baryon exploited here,
but the claim would be that these are the dominant components of the wave
function and they show up clearly in the electromagnetic properties studied
here. The slower experimental fall down with $Q^2$ could be an indication of the contribution of genuine quark components, along the lines of the work of \cite{Hyodo:2008xr} as we discussed in Sec. \ref{sec:model}.

The discrepancies found in the normalization of $A_{1/2}^p$ for the proton
deserve more attention. We have already commented  that should one use the 
width of the $N^*(1535)$ of 90 MeV of BES, or 100 MeV of MAID2007, the 
experimental
 values would be lowered and the agreement between theoretical results and the experiment would be better. In fact, the agreement of the theoretical results for $A_{1/2}^p$ with the MAID2007 analysis is very good, as we have already noticed. But then we could look at the ratio $R=S_{1/2}^p/A_{1/2}^p$ and we find $R=0.6$ at $Q^2=0.5\, {\rm GeV}^2$. Experimentally, this ratio is $R\sim 0.2$ if we take an average value of $S_{1/2}$ over the different data, so the discrepancies in this ratio seem to be large. Certainly, the experimental ratio becomes much larger if we take the points with open triangles in Fig. \ref{fig:S1/2proton}, and then $R\sim 0.44$. This large dispersion of experimental values is understandable if one recalls that the contributions of the $S_{1/2}$ term in the $ep\to e'p\eta$ cross section (from where the data is extracted) is of the order of a few percent \cite{burkert}. This, together with the experimental uncertainties in the normalization noted above, clearly indicate that large uncertainties in the experimental $S_{1/2}$ are indeed present. Further improvements in $S_{1/2}$ in the future will reveal if the discrepancies in the ratio $R$ pointed out here are deficiencies of the model or stem from present experimental uncertainties, or both. But it is clear that stronger claims in favor of the theoretical model are tied to a better precision in this experimental ratio, thus providing a justification for improved measurements of this magnitude.

  At the same time we addressed  the problem of eta photoproduction on the
 proton and neutron with the same formalism. We found a cross section 
 compatible with experiment in the $\gamma p \to \eta p$ reaction.
 This
 cross section also served to show evidence that our approach misses strength of the
 reaction at energies beyond the $N^*(1535)$. This could be due to the fact that
 the width that we obtain for the resonance, of the order of $75-90$ MeV, is
 smaller than the experimental one, or that the $\gamma p \to \eta p$ reaction
 collects strength from higher energy resonances which are not dynamically
 generated and hence do not appear in our scheme. This issue is not settled in
 view of the large dispersion of results that one can find in the literature for
 the width of the $N^*(1535)$, from about $90$ MeV  to $350$ MeV. Furthermore,
 the ratio of the cross sections of $\gamma n \to \eta n$ to
 $\gamma p \to \eta p$ was obtained in fair agreement with experiment,
 particularly at energies close to the $N^*(1535)$. 
 
     Altogether, the information extracted in this paper provides support
 for the idea of the  $N^*(1535)$ resonance as being largely dynamically generated from the
 interaction of mesons and baryons, the dynamics of which seems to be well
 accounted for  by chiral Lagrangians together with a proper coupled channels
 unitary treatment of the interaction, as provided by the chiral unitary
 approach.

{However, we also discussed that the recent study of \cite{Hyodo:2008xr} indicates the need for a genuine quark component of the $N^*(1535)$, which could provide strength at large $Q^2$ where our model, both in the nonrelativistic and relativistic versions, still provides a faster fall down with $Q^2$ than experiment.}

%\section*{Acknowledgments}  
\begin{acknowledgments}
D.J. wishes to acknowledge the hospitality of the University of Valencia, where part of this work was done. 
This work is partly supported by 
DGICYT contract number FIS2006-03438, 
the Generalitat Valenciana, 
the collaboration agreement between the JSPS of Japan and the CSIC of Spain,  
the Grant for Scientific Research (Nos.18042001, 20028004) 
and 
the Grant-in-Aid for the 21st Century COE ``Center for Diversity and Universality in Physics" from the Ministry of Education, Culture, Sports, Science and Technology of Japan.
This research is  part of
the EU Integrated Infrastructure Initiative  Hadron Physics Project under 
contract number RII3-CT-2004-506078,
and was done under Yukawa International Program for Quark-Hadron Sciences. 
\end{acknowledgments}

\end{document}